\newif\ifcmp\cmpfalse
\newif\ifthesis\thesisfalse

\ifcmp
\documentclass[cmp,draft,numbook]{svjour}
\else
\documentclass[12pt]{amsart}
\fi

\ifcmp
\usepackage{amsmath}
\usepackage{amsfonts,amssymb}
\else
\usepackage{amssymb}
\fi

\ifcmp
\journalname{Communications in Mathematical Physics}
\fi

\ifcmp
\spnewtheorem{teoreema}{Theorem}[section]{\bf}{\it}
\spnewtheorem{apulause}[teoreema]{Lemma}{\bf}{\it}
\spnewtheorem{korollaari}[teoreema]{Corollary}{\bf}{\it}
\spnewtheorem{propositio}[teoreema]{Proposition}{\bf}{\it}
\spnewtheorem{definitio}[teoreema]{Definition}{\bf}{\it}
\spnewtheorem{remarkki}[teoreema]{Remark}{\it}{\rm}
\spnewtheorem{esimerkki}[teoreema]{Example}{\it}{\rm}
\else
\theoremstyle{plain}
\newtheorem{teoreema}{Theorem}[section]
\newtheorem{apulause}[teoreema]{Lemma}
\newtheorem{korollaari}[teoreema]{Corollary}
\newtheorem{propositio}[teoreema]{Proposition}

\theoremstyle{definition}
\newtheorem{definitio}[teoreema]{Definition}

\theoremstyle{remark}
\newtheorem{remarkki}[teoreema]{Remark}

\fi

\numberwithin{equation}{section}

\newcommand{\thmref}[1]{Theorem~\ref{#1}}
\newcommand{\secref}[1]{\S\ref{#1}}
\newcommand{\lemref}[1]{Lemma~\ref{#1}}
\newcommand{\remref}[1]{Remark~\ref{#1}}
\newcommand{\defref}[1]{Definition~\ref{#1}}
\newcommand{\corref}[1]{Corollary~\ref{#1}}
\newcommand{\proref}[1]{Proposition~\ref{#1}}
\newcommand{\appref}[1]{Appendix~\ref{#1}}
\newcommand{\itemref}[1]{(\ref{#1})}
\newcommand{\equref}[1]{(\ref{#1})}

\newcommand{\commentout}[1]{}

\newcommand{\Mn}{M_n}
\newcommand{\R}{\mathbb{R}}
\newcommand{\Sph}{\mathbb{S}}
\newcommand{\N}{\mathbb{N}}
\newcommand{\Mat}{\mathbb{M}}
\newcommand{\prob}{\mathbb{P}}
\DeclareMathAlphabet{\mathfat}{U}{bbold}{m}{n}     
\newcommand{\one}{\mathfat{1}}

\DeclareMathOperator{\supp}{supp}
\DeclareMathOperator{\GL}{GL}

\DeclareMathOperator{\rk}{rk}
\DeclareMathOperator{\esup}{ess\ sup}
\DeclareMathOperator{\einf}{ess\ inf}
\DeclareMathOperator{\Dom}{Dom}
\DeclareMathOperator{\Dgn}{Dgn}
\DeclareMathOperator{\re}{Re}

\begin{document}

\title{Passive advection and the degenerate elliptic operators $\Mn$}

\author{Ville Hakulinen}

\ifcmp
\institute{Department of Mathematics, P.O.Box 4, 00014 University of Helsinki, Finland,\\ \email{Ville.Hakulinen@Helsinki.FI}}
\else
\address{Department of Mathematics, P.O.Box 4, 00014 University of Helsinki, 
Finland}
\email{Ville.Hakulinen@Helsinki.FI}
\fi

\ifcmp
\else
\date\today
\thanks{Typeset by \AmS-\LaTeX.}
\fi

\newcommand{\abstracttext}{
We prove estimates for the stationary state $n$-point functions
at zero molecular diffusivity in the Kraichnan model \cite{Kr}.
This is done by proving upper bounds for the heat kernels and
Green's functions of the degenerate elliptic operators $\Mn$
that occur in the Hopf equations for the $n$-point functions.
}

\newcommand{\acknowledgementstext}
{
The author would like to thank his thesis advisor Professor Antti Kupiainen
for his support, guidance and patience during the preparation of this article
and Professor N. Th. Varopoulos for teaching the argument used
in \secref{sec-local} in the uniformly elliptic case.
}

\ifcmp
\else
\begin{abstract}
\abstracttext
\end{abstract}
\fi

\maketitle

\commentout{
\section*{TODO List}
\begin{enumerate}
\end{enumerate}
}

\ifcmp
\abstract\abstracttext
\fi

\section{Introduction}
The Kraichnan model of passive advection is an exactly solvable
model that has a very similar phenomenology to the full Navier-Stokes
turbulence, but is much simpler in many respects. I'll only give a very
short reminder for the reader. More detailed introductions
to the problem we are addressing can be found e.g. in \cite{G} and \cite{K}.
See also \cite{EVE}, \cite{LJR1} and \cite{LJR2}.

Let $T(t,x)\in\R, x\in\R^d$ be a scalar quantity satisfying
\begin{equation}
\partial_t T=\kappa\Delta T-v\cdot\nabla T+f.
\end{equation}
In the Kraichnan model we take $v$ and $f$ random, decorrelated in time, independent
and Gaussian with mean zero and covariances

\begin{equation}\label{equ-covv}
\langle v^\alpha(t_1,x_1)v^\beta(t_2,x_2)\rangle = D^{\alpha\beta}(x_1-x_2)\delta(t_1-t_2)\text{ and}
\end{equation}
\begin{equation}\label{equ-covf}
\langle f(t_1,x_1)f(t_2,x_2)\rangle = C(x_1-x_2)\delta(t_1-t_2).
\end{equation}

Here the $v\cdot\nabla T$ should be interpreted in the Stratonovich
sense. The incompressibility of the velocity field $v$ is guaranteed by taking
\begin{equation}
D^{\alpha\beta}(x)=\int e^{-ik\cdot x}D(|k|)\biggl(\delta^{\alpha\beta}-\frac{k^\alpha k^\beta}{k^2}\biggr)\,dk
\end{equation}
where $D$ is smooth, nonnegative and of compact support in $(0,\infty)$. 
A $D$ that mimics turbulent velocities is
\begin{equation}
D(|k|)=|k|^{-(d+\xi)}\chi\biggl(|k|\eta+\frac{1}{|k|\ell}\biggr)
\end{equation}
with $\chi$ smooth, $\chi=1$ in a neighbourhood of the origin and $\chi(x)=0$ for $x>1$. The idea is that $D$ behaves like $|x|^\xi$ in the so-called
inertial range $\eta<<|x|<<\ell$. The number $\eta$ is called the Kolmogorov
scale and $\ell$ is called the inertial scale. We let
$\tilde{C}\in C^\infty_0(\R^d)$ with a nonnegative Fourier transform
and $C:=\tilde{C}(\cdot/L)$, with $L>0$.

One is interested in the statistics of $T(t,x)$ as $t\rightarrow\infty$.
Let
\begin{equation}
{\mathcal F}_n(t,x_1,...,x_n):=\langle T(t,x_1)...T(t,x_n)\rangle.
\end{equation}
Given \equref{equ-covv} and \equref{equ-covf} the $n$-point
functions ${\mathcal F}_n$ of the scalar $T$ obey the so-called Hopf equations
(see \cite{LJR2}):
\begin{multline}\label{equ-hopf}
\partial_t {\mathcal F}_n(t,x_1,...,x_n)=-{\mathcal M}_n {\mathcal F}_n
(t,x_1,...,x_n)+\\\sum_{1\le i<j\le n}{\mathcal F}_{n-2}(t,x_1,\underset{\hat{i}\hat{j}}{\ldots},x_{n})C(x_i-x_j),
\end{multline}
with
\begin{equation}\label{equ-mn}
\mathcal M_n:=-\sum_{1\le i<j\le n}\sum_{1\le\alpha,\beta\le d}
D^{\alpha\beta}(x_i-x_j)\frac{\partial^2}{\partial x^\alpha_i\partial x^\beta_j}-\kappa\sum_{1\le i\le n}\Delta_i.
\end{equation}

The fact that the Hopf equation for $\mathcal F_n$ does not contain
$\mathcal F_m$ with $m>n$ makes it easy to solve these equations
inductively. The situation here differs drastically from full Navier-Stokes
turbulence, where the Hopf equation for $\mathcal F_n$ contains also
$\mathcal F_{n+1}$.

$\mathcal M_n$ is an elliptic operator and in terms of its heat
kernel $\mathcal F_n$ (with zero initial condition for simplicity)
is given by
\begin{equation}
{\mathcal F}_2(t,\mathbf x)=\int_{t_0}^t ds\,\int d\mathbf y\,e^{-(t-s)\mathcal M_2}(\mathbf x,\mathbf y)C(y_1-y_2)
\end{equation}
\begin{multline}
{\mathcal F}_{2n}(t,\mathbf x)=\sum_{1\le i<j\le 2n}\int_{t_0}^t ds\,\int d\mathbf y\,e^{-(t-s)\mathcal M_{2n}}(\mathbf{x},\mathbf{y})\cdot\\
\cdot\mathcal{F}_{2n-2}(s,y_1,\underset{\hat{i}\hat{j}}{\ldots},y_{2n})C(y_i-y_j)\,d\mathbf y
\end{multline}
with vanishing odd correlators.

As $t_0\rightarrow -\infty$ these have the stationary limit
\begin{equation}
{\mathcal F}_2=\int d{\mathbf y}\,({\mathcal M}_2)^{-1}(\mathbf{x},\mathbf{y})C(y_1-y_2)
\end{equation}
\begin{equation}
{\mathcal F}_{2n}=\sum_{1\le i<j\le 2n}\int({\mathcal M}_{2n})^{-1}(\mathbf{x},\mathbf{y})
\mathcal{F}_{2n-2}(y_1,\underset{\hat{i}\hat{j}}{\ldots},y_{2n})C(y_i-y_j)\,d\mathbf y.
\end{equation}

One is interested in the study of $\mathcal F_{2n}$ for $\eta$ small,
$\ell$ large, $\kappa$ small and $L$ large. In this paper we prove
bounds for these directly in the limit $\eta=0$, $\ell=\infty$ and
$\kappa=0$ with fixed $L$, say $L=1$. Our methods also allow the study
of the limit $\eta\rightarrow 0$, $\ell\rightarrow\infty$ and
$\kappa\rightarrow 0$ \cite{H}.

A comment on $D$ is now in place. While sending $\eta\rightarrow 0$
and $\ell\rightarrow\infty$ in $D$, we get into trouble with $\ell$, since $D$
diverges as $\ell\rightarrow\infty$. Fortunately it doesn't matter: Let
\begin{equation}
d^{\alpha\beta}(x):=\int dk\,(1-e^{ik\cdot x})D(|k|)\biggl(\delta^{\alpha\beta}
-\frac{k^\alpha k^\beta}{k^2}\biggr).
\end{equation}

Now \equref{equ-mn} can be written in the following form:
\begin{equation}
\mathcal M_n:=\sum_{1\le i<j\le n}\sum_{1\le\alpha,\beta\le d}
d^{\alpha\beta}(x_i-x_j)\frac{\partial^2}{\partial x^\alpha_i\partial x^\beta_j}-\kappa\Delta-D_0\biggl(\sum_{1\le i\le n}\sum_{1\le\alpha\le d}\frac\partial{\partial_{x^\alpha_i}}\biggr)^2
\end{equation}

In \equref{equ-hopf} $\mathcal M_n$ acts on translationally invariant functions,
so the last term drops out and the rest has a limit as $\ell\rightarrow\infty$.

Finally, here's our main Theorem, proved directly at $\eta=0$, $\ell=\infty$
and $\kappa=0$:
\begin{teoreema}\label{thm-main2}
\begin{equation}
\mathcal{F}_{2n}(\mathbf x)\le C_n\sum_{\pi}\prod_{\{i,j\}\in\pi}(1+|x_i-x_j|)^{2-\xi-d},
\end{equation}
where the sum is over pairings of $\{1,...,2n\}$.
\end{teoreema}

\section{Preliminaries}
This section fixes the notation and discusses the results from other
papers (\cite{CW}, \cite{D}, \cite{GW}, \cite{V}) used in this paper. There is an overview
of this paper in \secref{sec-ovw}, so the reader might want to start
there.

\subsection{Degenerate elliptic operators in divergence form}

Let $\Omega\subset \R^n$ be a domain. We shall be interested
in second order differential operators in divergence form, i.e.
in operators $H$ of the form $H=-\nabla\cdot A\nabla$, where
$A$ is a locally square integrable function from $\Omega$ to real symmetric
positive $n\times n$ matrices with locally square integrable distributional
derivative, i.e. $A\in W^{1,2}_{\mathrm{loc}}(\Omega,\Mat^n)$.
One can make sense of more general operators, but
this is not relevant to the results presented in this paper.

\begin{definitio}
Let $H$ and $A$ be as above. The matrix $A$ is called the \emph{symbol} of $H$,
and we denote $\sigma(H):=A$.
The function $w^H_1(x):=\inf_{v\in\Sph^{n-1}}\langle v,\sigma(H)(x)v\rangle$ (resp.
$w^H_2(x):=\sup_{v\in\Sph^{n-1}}\langle v,\sigma(H)(x)v\rangle$)
is called the greatest lower bound (resp. least upper bound) of the symbol.
\end{definitio}

We shall also use $\sigma(H)$ to denote the quadratic form
$\langle v, A(x)v\rangle$. The usage will be clear from the context.
We often speak loosely and forget the attributes ``greatest'' and ``lowest''
from the bounds.

If $A$ and $B$ are two symbols and $U\subseteq\R^m$, we shall denote
$A\sim^\lambda B$ on $U$, if 
$\lambda A\le B \le \lambda^{-1}A$ a.e. on $U$.
If there is $\lambda>0$ so that $A\sim^\lambda B$ on $U$ we also
say $A\sim B$ on $U$. If ``on $U$'' is dropped, we refer to whole
$\R^m$.

We shall use $\one$ to denote the identity matrix. Thus a symbol $A$
is uniformly elliptic iff $A\sim \one$. Moreover, if $A$ and $B$
are symbols on $\R^{n_1}$ and $\R^{n_2}$, then $A\oplus B$
is just the natural symbol on $\R^{n_1+n_2}$.

\begin{definitio}\label{def-ats}
Let $w$ be a nonnegative locally integrable function (a weight)
defined on $\R^n$. We
denote $w(A):=\int_A w(x)dx$. The function $w$ is called a
\emph{doubling weight} (resp. an \emph{$A_2$-weight}), if there is
a constant $C$ such that for every ball $B\subset\R^n$ we have
$w(2B)\le Cw(B)$ (resp. 
$\frac{1}{|B|^2}w(B)w^{-1}(B)\le C$).
\end{definitio}

Since by Schwartz inequality $|B|^2\le w(B)w^{-1}(B)$, we have
$|2B|^2=2^{2n}|B|^2\le 2^{2n}w(B)w^{-1}(B)\le 2^{2n}w(B)w^{-1}(2B)$,
so we can conclude that an $A_2$-weight is also a
doubling weight.

\begin{definitio}\label{def-poin}
Denote $u_B:=|B|^{-1}\int_Bu(x)\,dx$ and let $w_1,w_2$ be weights on
$\R^n$ and let $q>2$. We say that the \emph{Poincar\'e inequality} 
(resp. \emph{Sobolev inequality}) holds
for $w_1$, $w_2$ with $q$, if there is $C<\infty$ so that for every
ball $B\subseteq\R^n$ and $u\in W^{1,2}(B)$ (resp. $u\in W^{1,2}_0(B)$) we have
\begin{equation}
\begin{split}
\biggl(w_2(B)^{-1}\int_B|u-u_B|^q w_2&\,dx\biggr)^{1/q}\\
&\le C|B|^{1/n}\biggl(w_1(B)^{-1}\int_B|\nabla u|^2 w_1\,dx\biggr)^{1/2}
\end{split}
\end{equation}
(resp.
\begin{equation}
\begin{split}
\biggl(w_2(B)^{-1}\int_B|u|^q w_2&\,dx\biggr)^{1/q}\\
&\le C|B|^{1/n}\biggl(w_1(B)^{-1}\int_B|\nabla u|^2 w_1\,dx\biggr)^{1/2}
\end{split}
\end{equation}
).
\end{definitio}

\begin{teoreema}\label{thm-har}
\emph{(Harnack inequality)} Suppose $H:=-\nabla\cdot A\nabla$ is a divergence
form operator with $w_1\le A \le w_2$ and suppose that the weights $w_1$
and $w_2$ satisfy the following:
\begin{enumerate}
\item $w_1$ and $w_2$ are in $A_2$,
\item The Poincar\'e inequality holds for $w_1$, $w_2$ with some $q>2$ and
\item The Poincar\'e inequality holds for $w_1$, $1$ with some $q'>2$.
\end{enumerate}
Let $t_0,...,t_4\in\R$ with $t_0<...<t_4$, $\Omega\subseteq\R^n$ open and
$K\subseteq\Omega$ compact and connected. Let $u$ be a strictly positive 
solution to $u_t+Hu=0$ in $\Omega\times(t_0,t_4)$. Then there is a
constant $C<\infty$ depending on $\Omega$, $K$ and $t_0,...,t_4$, but
on $A$ only through the bounds $w_1$ and $w_2$ so that
\begin{equation}
\esup_{K\times(t_1,t_2)}u\le C\einf_{K\times(t_3,t_4)}u
\end{equation}
\end{teoreema}

\begin{proof}
This is just Theorem A of \cite{GW} supplemented with a covering argument
from \cite{M}, pages 734-736.\ifcmp\qed\fi
\end{proof}

\begin{remarkki}\label{rem-solution}
For the purposes of \thmref{thm-har} the concept of $u$ being a
solution of $u_t+Hu=0$ on $Q:=\Omega\times(t_0,t_4)$ means exactly the
following:
\begin{enumerate}
\item \label{item-uinltwo} $u\in L^2(Q)$,
\item \label{item-utinltwo} $u_t\in L^2(Q)$,
\item \label{item-uginltwo} $|\nabla u|^2w_2\in L^1(Q)$ and
\item \label{item-usol} For all $\phi\in C^1_0(Q)$
we have 
\begin{equation}
\int_Q u_t\phi+\langle A\nabla u,\nabla\phi\rangle\,dx\,dt=0
\end{equation}
\end{enumerate}

We are going to apply to apply the Harnack inequality only to heat
kernels of some degenerate elliptic operators. In particular as
long as $t_0>0$ all the above items will hold. 

Since the heat kernel is a positive distribution, it is a measure and
\itemref{item-usol} follows from the fact that the heat kernel is
a distributional solution of the corresponding degenerate heat equation.

First of all \itemref{item-uinltwo} holds because for $t_0>0$ the heat kernel
is a bounded function on $\Omega\times(t_0,t_4)$ (by \corref{cor-dim}).

Secondly \itemref{item-utinltwo} holds because of the following
computation which is justified by \remref{rem-strocon}:
\begin{equation}
\begin{split}
(\partial_tK)(s,\cdot,y)&=-HK(s,\cdot,y)\\
&=-e^{-(s-t_0)H}He^{-t_0H/2}K(\frac{t_0}2,\cdot,y).
\end{split}
\end{equation}
Now since by \remref{rem-contr} $e^{-tH}$ is a contraction on $L^2$,
\begin{equation}
\sup_{s\in(t_0,t_4)}||\partial_t K(s,\cdot,y)||_2<\infty.
\end{equation}

Let $A$ be the symbol of $H$.
To prove \itemref{item-uginltwo} it suffices to show that $|\nabla K|$
is locally in $L^2$, since $w_2$ is locally bounded. Since
\begin{equation}
\int_Q |\nabla K|^2\,dx\,dt\le\int_Q w_1^{-1}\langle A\nabla K,\nabla K\rangle
\,dx\,dt.
\end{equation}

Since $w_1$ is in $A_2$ (by \lemref{lem-beh}), $w_1^{-1}$ is locally
integrable, so it suffices to prove that
$\langle A\nabla K,\nabla K\rangle$ is essentially bounded on $Q$.
We show that for any $0\le\phi\in C^{\infty}_0(Q)$ we have
\begin{equation}
\int_Q \phi\langle A\nabla K,\nabla K\rangle\le C\int_Q\phi,
\end{equation}
with $C$ not depending on $\phi$.

So we compute using the facts that $K$ and $\nabla\cdot A\nabla K$
are locally bounded:
\begin{equation}
\begin{split}
\int_Q \phi&\langle A\nabla K,\nabla K\rangle=\biggl|\int_Q K\nabla\cdot\phi A\nabla K\biggr|\\
&\le C\biggl|\int_Q \langle A\nabla \phi,\nabla K\rangle\biggr|+C\biggl|\int_Q
\phi\nabla\cdot A\nabla K\biggr|\\
&\le 2C\biggl|\int_Q \phi \nabla\cdot A\nabla K\biggr|\le C'\int_Q\phi.
\end{split}
\end{equation}
\end{remarkki}

It follows from the results in \secref{sec-bnd} and \appref{app-poin}
that this Harnack inequality holds for the operators $M_n$, which
will be our main interest and will be defined in \secref{sec-def-mn}.

\subsection{Gaussian upper bounds for heat kernels}

The material in this section is mostly taken from \cite{D}. For more
information, see sections 1.3, 2.4 and 3.2 there. See also \cite{DA} and
\cite{V}.

\begin{definitio}
Let $H\ge 0$ be a real self-adjoint operator on $L^2(\R^n)$. We call
the semigroup $e^{-Ht}$ a \emph{symmetric Markov semigroup}, if
it is positivity-preserving and a contraction on $L^\infty(\R^n)$.
\end{definitio}

\begin{remarkki}\label{rem-contr}
By saying that $e^{-Ht}$ is a contraction on $L^p$ with $p\not=2$
we mean that $e^{-Ht}$ is a contraction on $L^p\cap L^2$ and can be extended
to a unique contraction on $L^p$. In the case of $L^\infty$ we have
to impose the extra condition of weak* continuity to achieve uniqueness
since $L^\infty\cap L^2$ is not norm dense in $L^\infty$.
\end{remarkki}

\begin{remarkki}\label{rem-strocon}
A symmetric Markov semigroup is strongly continuous on $L^p$ with
$1\le p<\infty$, see Theorem 1.4.1 of \cite{D}. This in particular
implies that the generator $H$ commutes with the semigroup $e^{-Ht}$
(see \cite{DA}).
\end{remarkki}

By Theorem 1.3.5 of \cite{D}, any self-adjoint divergence form operator
with non-negative symbol and core $C^\infty_0(\R^n)$ gives rise to a
symmetric Markov semigroup. The
Theorem there is stated for ``elliptic'' operators, but the proof works
for any non-negative symbol. The keywords here are self-adjointness
and core $C^\infty_0$. Both follow for $M_n$ from the fact that
$\sigma(M_n)\in W^{1,2}_{\mathrm{loc}}(\R^{(n-1)d})$ (\proref{pro-squ}).
See Theorem 1.2.5 of \cite{D}.

\begin{definitio}
Let $e^{-Ht}$ be a symmetric Markov semigroup on $L^2(\R^n)$. We say
that $e^{-Ht}$ is \emph{ultracontractive} if the map $e^{-Ht}$
is bounded from $L^2$ to $L^\infty$ for every $t>0$.
\end{definitio}

\begin{definitio}\label{def-dim}
Suppose that $C^\infty_0(\R^n)\subseteq \Dom(H)$. Let $e^{-Ht}$ be a symmetric
Markov semigroup on $L^2(\R^n)$. We say that $e^{-Ht}$ (or $H$ or $\sigma(H)$)
is of \emph{dimension} $\mu$ if there is
$C_2<\infty$ such that for all $t>0$ and
$f\in L^2(\R^n)$ we have
\begin{equation}
||e^{-Ht}f||_\infty\le C_2 t^{-\mu/4}||f||_2.
\end{equation}
\end{definitio}

Note that the dimension of a semigroup need not be unique.

There is a standard method for obtaining global Gaussian upper bounds
for heat kernels of divergence form operators with nonnegative symbols
using global space-independent bounds. A good reference for this is \cite{DA}.

\begin{definitio}\label{def-met}
Let $A$ be a symbol on $\R^n$. The function
\begin{equation}
\begin{split}
d_A(x,y):=&\sup\bigl\{|\phi(x)-\phi(y)|:\phi\text{ is }C^\infty\text{ and
bounded with }\\
        &\langle\nabla\phi,A\nabla\phi\rangle\le 1\text{ on }\R^n\bigr\}
\end{split}
\end{equation}
is called the metric associated with $A$
(or $H$, if $H:=-\nabla\cdot A\nabla$ or $e^{-tH}$ or the heat kernel
of $H$).
\end{definitio}

The following Theorem was proved by Davies \cite{DA}.
\begin{teoreema}\label{thm-gub}
Let $\mu$ be a positive real number. Suppose
$H:=-\nabla\cdot A\nabla\ge 0$ is a positive self-adjoint divergence
form operator with $e^{-Ht}$ a symmetric Markov semigroup of dimension $\mu$.
Then for each $\delta>0$ there is $C_\delta<\infty$ such that
the heat kernel $K$ of $e^{-Ht}$ satisfies
\begin{equation}
0\le K(t,x,y)\le C_\delta t^{-\mu/2}\exp\{-\frac{d_A(x,y)^2}{4(1+\delta)t}\}
\end{equation}
for all $0<t<\infty$ and $x,y\in\R^n$. Besides $\delta$, $C_\delta$
depends only on $\mu$ and the constant $C_2$ of \defref{def-dim}.
\end{teoreema}

\begin{proof}
See \cite{DA}.\ifcmp\qed\fi
\end{proof}

We shall use the following Theorem later to get the dimension of $M_n$
in \corref{cor-dim}. It was proved by Varopoulos \cite{V}. John Nash \cite{N} 
also proved a similar result.

\begin{teoreema}\label{thm-sou}
Suppose $C^\infty_0(\R^n)\subseteq \Dom(H)$. Let $e^{-Ht}$ be a symmetric Markov
 semigroup on $L^2(\R^n)$ and let
$\mu > 2$ be given. Then there is $C_1<\infty$ such that for all
$f\in C^\infty_0(\R^n)$ we have
\begin{equation}
||f||^2_{2\mu/(\mu-2)}\le C_1 \langle f,Hf \rangle.
\end{equation}
if and only if there is $C_2<\infty$ such that for for all $t>0$ and
$f\in L^2(\R^n)$ we have
\begin{equation}\label{equ-ultracontr}
||e^{-Ht}f||_\infty\le C_2 t^{-\mu/4}||f||_2.
\end{equation}
Here the constants $C_1$ and $C_2$ depend only on each other and the number $\mu$.
\end{teoreema}

\begin{proof}
This is just Theorem 2.4.2 of \cite{D}.\ifcmp\qed\fi
\end{proof}

\begin{remarkki}\label{rem-kernel}
One can show using the Schwartz Kernel and Radon-Nikodym Theorems
that a bounded linear map $L:L_1\rightarrow L_\infty$ has a integral
kernel that is a function in $L_\infty$ whose $L_\infty$-norm
equals the operator norm of $L$. Since our $e^{-Ht}$ is self-adjoint,
boundedness of $e^{-Ht}:L_2\rightarrow L_\infty$ implies boundedness
of $e^{-Ht}:L_1\rightarrow L_2$, so in this case we have a heat
kernel that is a genuine function.
\end{remarkki}

Finally, we give a nice way to estimate heat kernels of operators $H$
that have symbols satisfying $\sigma(H)\sim A_1\oplus A_2$.

\begin{teoreema}\label{thm-prod}
Suppose that for $i=1,2$, $A_i$ is a symbol on $\R^{n_i}$ such that
$e^{t\nabla\cdot A_i\nabla}$ is a symmetric Markov semigroup on $L^2(\R^{n_i})$
and $B\sim^\lambda A_1\oplus A_2$. Suppose also that the heat kernels of
$A_i$'s satisfy
\begin{equation}
K_{A_i}(t,x,y)\le C_i t^{-\frac{\mu_i}2} \exp\{-\frac{d_{A_i}(x,y)^2}{C_i t}\}.
\end{equation}
Then there is $C<\infty$ depending only on $C_1$, $C_2$, $\mu_1$, $\mu_2$
and $\lambda$ so that the heat kernel of $B$ satisfies
\begin{equation}
K_B(t,x,y)\le C t^{-\frac{\mu_1+\mu_2}2} \exp\{-\frac{d_{A_1}(x,y)^2+d_{A_2}(x,y)^2}{Ct}\}.
\end{equation}
\end{teoreema}

\begin{proof}
Since $K_{A_1\oplus A_2}(t,(x_1,x_2),(y_1,y_2))=K_{A_1}(t,x_1,y_1)K_{A_2}(t,x_2,y_2)$, we can conclude that
\begin{equation}
K_{A_1\oplus A_2}\le C_1 C_2 t^{-\frac{\mu_1+\mu_2}2},
\end{equation}
which by Riesz-Thorin interpolation theorem and the fact that
$e^{t\nabla\cdot A_1\oplus A_2\nabla}$ is a contraction $L^\infty$
imply \equref{equ-ultracontr} for $H=-\nabla\cdot A_1\oplus A_2\nabla$.
Therefore by \thmref{thm-sou}
\begin{equation}
||f||^2_{2\mu/(\mu-2)}\le C_3 \langle \nabla f,(A_1\oplus A_2)\nabla f \rangle
\end{equation}
for any $f\in C^\infty_0(\R^{n_1+n_2})$ with $C_3$ depending only on
$C_1 C_2$ and $\mu_1+\mu_2$. Since $A_1\oplus A_2\le \lambda^{-1} B$,
we have
\begin{equation}
K_B\le C_4 t^{-\frac{\mu_1+\mu_2}2},
\end{equation}
with $C_4$ depending only on $C_1 C_2$, $\mu_1+\mu_2$ and $\lambda$.
We now apply \thmref{thm-gub} to conclude the claim.\ifcmp\qed\fi
\end{proof}

\subsection{The definition of the operators $M_n$}\label{sec-def-mn}
For the rest of the paper, we fix a constant $\xi$, $0<\xi<2$ and
an integer $d\ge 2$. Here $d$ is the dimension of the ``physical'' space.

Next, we overload the symbol $d$ immediately and let $d$ be the map
from $R^d$ to $d\times d$ matrices defined by
\begin{equation}
d(x):=C \int_{\R^d}\frac{1-\cos(k\cdot x)}{|k|^{d+\xi}}(1-\hat k\otimes\hat k)\,dk,
\end{equation}
with
\begin{equation}
C:=\frac{(4\pi)^{d/2}2^\xi \xi \Gamma((d+\xi+2)/2)}{(d-1)\Gamma((2-\xi)/2)}.
\end{equation}

A computation (see e.g. \cite{EX}) shows that
\begin{equation}
d(x)=|x|^\xi\biggl((1+\frac{\xi}{d-1})\one-\frac{\xi}{d-1}\hat{x}\otimes\hat{x}\biggr).
\end{equation}

In the following definition,
we denote vectors in $\R^{nd}$ by $\{v_i\}_{i=1}^n$, where
each $v_i$ is a vector in $\R^d$.

\begin{definitio}
Let $n\ge 2$. The operator
$\mathcal M^{sc}_n:=-\nabla\cdot\sigma(\mathcal M^{sc}_n)\nabla$
is the one with the symbol
\begin{equation}
\sigma(\mathcal M^{sc}_n):=-\sum_{1\le i < j \le n} \langle v_i, d(x_i-x_j)v_j\rangle
\end{equation}
\end{definitio}

If $a\in\R^d$, we denote the vector $(x_i+a)_{i=1}^n$ by $\mathbf x+a$.
We call a function $f:\R^{nd}\rightarrow
\R$ translationally invariant, if for every $a\in \R^d$ and
$\mathbf x\in\R^{nd}$ we have $f(\mathbf x)=f(\mathbf x+a)$.

We shall be interested in $\mathcal M^{sc}_n$ acting on translationally
invariant functions, so we need to reduce the number of total space
dimensions to $(n-1)d$.

In other words, we set $x_i:=y_i-y_{i+1}$ for $1\le i \le n-1$, so
\begin{equation}
\frac{\partial}
{\partial y^\alpha_i}=
\begin{cases}
\frac{\partial}{\partial x^\alpha_1}&\text{if $i=1$,}\\
\frac{\partial}{\partial x^\alpha_i}-\frac{\partial}{\partial x^\alpha_{i-1}}&\text{if $2\le i \le n-1$ and}\\
-\frac{\partial}{\partial x^\alpha_{n-1}}&\text{if $i=n$.}\\
\end{cases}
\end{equation}

Denote the symbol obtained in this way by $\sigma(\Mn)$.
A simple calculation shows that $\sigma(\Mn)$ equals
\begin{equation}
\sum_{i=1}^{n-1}\sum_{j=i}^{n-1}\langle v_i,(d(\sum_{k=i}^{j}x_k)-
d(\sum_{k=i}^{j-1}x_k)-d(\sum_{k=i+1}^{j}x_k)+d(\sum_{k=i+1}^{j-1}x_k))
v_{j}\rangle
\end{equation}
In particular,
\begin{equation}
\sigma(M_2)=\langle v_1,d(x_1)v_1\rangle,
\end{equation}
\begin{equation}
\begin{split}
\sigma(M_3)=&\langle v_1,d(x_1)v_1\rangle+\langle v_2,d(x_2)v_2\rangle+\\
            &\langle v_1,(d(x_1+x_2)-d(x_1)-d(x_2))v_2\rangle
\end{split}
\end{equation}
and
\begin{equation}
\begin{split}
\sigma(M_4)=&\langle v_1,d(x_1)v_1\rangle+\langle v_2,d(x_2)v_2\rangle+
                \langle v_3,d(x_3)v_3\rangle\\
            &\langle v_1,(d(x_1+x_2)-d(x_1)-d(x_2))v_2\rangle\\
            &\langle v_2,(d(x_2+x_3)-d(x_2)-d(x_3))v_3\rangle\\
            &\langle v_1,(d(x_1+x_2+x_3)-d(x_1+x_2)-d(x_2+x_3)+d(x_2))v_3\rangle
\end{split}
\end{equation}

\section{Overview}\label{sec-ovw}
Our intent here is to give some intuition on the arguments of this
paper and how they lead to the proof of \thmref{thm-main2}. What is
obvious at first sight, is that if \thmref{thm-main2} is to hold,
the Green's functions of the operators $M_{2n}$ should be
locally integrable in the sense that for every $n\ge 2$
there is $C<\infty$ so that for every $x\in\R^{(2n-1)d}$ we have
\begin{equation}\label{equ-loi}
\int_{B(x,1)}d^{(2n-1)d}y\,G_{M_{2n}}(x,y)<C.
\end{equation}

One might hope to get \equref{equ-loi} to hold using the 
heat kernel estimate of \thmref{thm-gub}, but unfortunately 
this direct approach fails. First of all we see
that $\sigma(M_2)\sim |\cdot|^\xi$. Applying \defref{def-met},
\corref{cor-dim} and \thmref{thm-gub} to this, we find a
$C<\infty$ such that
\begin{equation}\label{equ-est2}
K_{M_2}(t,x,y)\le C t^{-\frac d{2-\xi}}\exp\{-\frac{|x-y|^2}{Ct}\}
\end{equation}
for $|x|=1$ and $|x-y|\le \frac 12$. Integrating with respect to
$t$ from $0$ to $\infty$ we get
\begin{equation}
G_{M_2}(x,y)\le C'|x-y|^{2-\frac{2d}{2-\xi}}.
\end{equation}

This estimate yields \equref{equ-loi} only when $2-\frac{2d}{2-\xi}>-d$,
that is $\xi<\frac{4}{d+2}$. We might be satisfied with the fact that
\equref{equ-loi} holds only for small $\xi$, but there is worse to come:
For each $\sigma(M_n)$ will have points $x\in\Sph^{(n-1)d-1}$
so that $\sigma(M_n)\sim \one$ in a neighbourhood $U$ of $x$. A similar
argument as above now yields
\begin{equation}
G_{M_{2n}}(x,y)\le C'|x-y|^{2-\frac{2(n-1)d}{2-\xi}}
\end{equation}
for $y\in U$. This yields \equref{equ-loi} for $M_n$ only when
$\xi<\frac{4}{(n-1)d+2}$, which means trouble: Given $\xi$ with
$0<\xi<2$, there will always some be $N$ so that our argument above
fails to give local integrability for $M_n$ with $n\ge N$.

On the other hand, since $M_2$ is uniformly elliptic in a
neighbourhood $U$ of $x$, the heat kernel of $M_2$ should behave
like the heat kernel of the Laplacian for small times and small
distances from $x$.

Turning this analysis into formulas let's suppose
\begin{equation}\label{equ-est3}
K_{M_2}(t,x,y)\le C_2 t^{-\frac d2}\exp\{-\frac{|x-y|^2}{C_2 t}\}
\end{equation}
for $|x|=1$, $|x-y|\le\epsilon\le\frac 12$ and $0<t\le t_0$. Since
there is $C_3<\infty$ so that $t^{-\frac{d}{2-\xi}}\le C_3 t^{-\frac d2}$
for $t\ge t_0$, we can combine \equref{equ-est2} with
\equref{equ-est3} and conclude that
\begin{equation}\label{equ-est4}
K_{M_2}(t,x,y)\le C_4 t^{-\frac d2}\exp\{-\frac{|x-y|^2}{C_4 t}\}
\end{equation}
for $|x|=1$, $|x-y|\le\epsilon$ and $0<t<\infty$. Now
an integration w.r.t. $t$ from $0$ to $\infty$ yields
\begin{equation}
G_{M_2}(x,y)\le C_5 |x-y|^{2-d}
\end{equation}
for $|x|=1$ and $|x-y|\le\epsilon$. The same holds for $M_n$ with
$n>2$. This leads us to a further twist: for $n>2$, $\sigma(M_n)$
has degeneracies also outside of the origin, but fortunately in the end these
turn out not to be problematic.

A few words on the structure of the rest of the paper.
In \secref{sec_mn} the symbols of $M_n$ are analyzed in detail.
The local analysis of the heat kernels is done in \secref{sec-local}.
\thmref{thm-main2} is proved in \secref{sec-stat}.
Finally, there are three appendices containing technicalities.

\section{The operators $\Mn$}\label{sec_mn}
\ifthesis
 \pagestyle{myheadings}
  \markboth{CHAPTER \ref{chap-art}.\ \ THE ARTICLE\ %
    }%
    {\ref{sec_mn}.\ \ THE OPERATORS $\Mn$}
\fi

From now on, we live in $\R^{(n-1)d}$ and denote vectors of $\R^{(n-1)d}$ with
$\mathbf v=(v_i)_{i=1}^{n-1}$ and $\mathbf x=(x_i)_{i=1}^{n-1}$,
where $v_i,x_i\in \R^d$. 

The symbol of $\Mn$ has a bunch of useful symmetries, inherited from
$\mathcal M^{sc}_n$. For $L:\R^k\rightarrow\R^l$ a surjective linear
mapping and $A$ a symbol on $\R^k$ which for all $x\in\R^k$ is constant
on $\{x\}+\ker L$ denote $A^L(x):=L A(L^{-1}x) L^T$,
where $L^{-1}$ is some right-inverse of $L$. Let
$L_n:\R^{nd}\rightarrow\R^{(n-1)d}$ be given by the matrix
$(L_n)_{ij}:=\delta_{ij}-\delta_{i+d,j}$, so that
$\sigma(M_n)=\sigma(\mathcal M^{sc})^{L_n}$

We let
\begin{equation}
\mathcal L_n=\{L_n L L^{-1}_n:
L\text{ is a permutation of the coordinate axes of }\R^{nd}\}.
\end{equation}
Now $\sigma(M_n)^L=\sigma(M_n)$ for every $L\in\mathcal L_n$

\begin{remarkki}\label{rem-sym}
Let $A_1$ and $A_2$ be two symbols on $\R^k$ and let $G_1,G_2\subseteq
\GL(\R^k)$ be their respective symmmetry groups, i.e
\begin{equation}
G_i:=\{L\in\GL(\R^k):A_i^L=A_i\},
\end{equation}
for $i\in\{1,2\}$. Now if $A_1\sim A_2$ on
$U$, then $A_1\sim A_2$ on $LU$ for any $L\in G_1\cap G_2$.
\end{remarkki}

\begin{remarkki}\label{rem-deg}
A simple calculation shows that $\Mn$ is degenerate, whenever
\ifcmp{\linebreak[4]}\else{\linebreak[4]}\fi
$\sum_{i=a}^b x_i=0$, where $1\le a \le b \le n-1$. In fact these are the
only points where $\Mn$ degenerates, as we see in \thmref{thm-deg}.
To avoid lengthy statements in the rest of the paper,
we denote $\{\mathbf x\in \R^{(n-1)d}:x_i=0\}$ by $\{x_i=0\}$ and
similarly for the other sets.
\end{remarkki}

\begin{propositio}\label{pro-squ}
\begin{equation}
\sigma(M_n)\in W^{1,2}_{\mathrm{loc}}(\R^{(n-1)d})
\end{equation}
\end{propositio}

\begin{proof}
The case $1<\xi<2$ is an easy computation, since then $\sigma(M_n)$ is continuously differentiable.

In case $0<\xi\le 1$, we let 
\begin{equation}
F:=\bigcup_{1\le a \le b < n}\{\sum_{i=a}^b x_i=0\}.
\end{equation}
A relatively simple calculation shows that there is $C<\infty$ such that
\begin{equation}
|\nabla(\sigma(M_n))(\mathbf x)|\le C d(\mathbf x,F)^{\xi-1}.
\end{equation}

Since $F$ is a finite union of vector subspaces of codimension $d\ge 2$, 
we can conclude that $d(\mathbf x,F)^{\xi-1}$ is a locally square integrable
function.\ifcmp\qed\fi
\end{proof}

\begin{remarkki}\label{rem-upb}
It is trivial to get an upper bound for $\Mn$:
\begin{equation}
\sigma(M_n)\le (\sup_{|\mathbf y|=|\mathbf w|=1}\langle \mathbf w,\sigma(M_n)(\mathbf y)\mathbf w\rangle)|\mathbf x|^\xi |\mathbf v|^2.
\end{equation}
\end{remarkki}

We obtain a better upper bound in section \secref{sec-bnd}.

\begin{propositio}\label{pro-met}
For any $\epsilon\in(0,1)$ there is $C<\infty$ such that
\begin{equation}
d_{\sigma(M_n)}(x,y)\le C|x-y|^{1-\xi/2},
\end{equation}
when $|x-y|\ge \epsilon|x|$.
\end{propositio}

\begin{proof}
By \defref{def-met} and \remref{rem-upb} it suffices to show that
there is $C<\infty$ such that $d_{|\cdot|^\xi}(x,y)^2\le C|x-y|^{2-\xi}$,
when $|x-y|\ge \epsilon|x|$. Trivial dimensional analysis
gives $d_{|\cdot|^\xi}(x,y)= |x|^{1-\xi/2}d_{|\cdot|^\xi}(\hat x,\frac{y}{|x|})$. Therefore we may assume $|x|=1$. By rotational symmetry, we may fix
$x$. By scaling, there is $C'<\infty$ so that
$C'|y|^{1-\xi/2} = d_{|\cdot|^\xi}(0,y)$. Since now
\begin{equation}
\begin{split}
\frac{d_{|\cdot|^\xi}(x,y)^2}{|x-y|^{2-\xi}}&=
C'\frac{d_{|\cdot|^\xi}(x,y)^2}{d_{|\cdot|^\xi}(0,x-y)^2},
\end{split}
\end{equation}
it suffices to show that
$f(R):=\sup_{|x-y|=R}d_{|\cdot|^\xi}(x,y)/d_{|\cdot|^\xi}(0,x-y)$ is a bounded
function of $R$ for $R\in[\epsilon,\infty)$. Obviously $f$ is continuous.
By continuity of $d_{|\cdot|^\xi}$ we have
\begin{equation}
\frac{d_{|\cdot|^\xi}(x,y)^2}{d_{|\cdot|^\xi}(0,x-y)^2}=
\frac{d_{|\cdot|^\xi}(\frac{x}{|x-y|},\frac{y}{|x-y|})^2}
{d_{|\cdot|^\xi}(0,\widehat{x-y})^2}\rightarrow
\frac{d_{|\cdot|^\xi}(0,\hat{y})^2}{d_{|\cdot|^\xi}(0,\hat{y})^2}=1
\end{equation}
as $|x-y|\rightarrow\infty$.
\ifcmp\qed\fi
\end{proof}

\subsection{Fourier integral representation and the degeneration set}
\begin{definitio}
Let $A$ be a symbol. We call the set
\begin{equation}
\Dgn(A):=\{x\in R^n:A(x)\text{ is not invertible}\}
\end{equation}
the degeneration set of $A$.
\end{definitio}

The following Fourier integral representation of the symbol
is crucial for the computation of the degeneration sets of
$\mathcal M_n$ (which then implies corresponding properties
for the operators $\Mn$ to be introduced later).

\begin{teoreema}\label{thm-deg}
The degeneration set of $\Mn$ is
\begin{equation}
\Dgn(\Mn)=\bigcup_{1\le i \le j < n}\{\mathbf x\in R^{(n-1)d}:|x_i+...+x_j|=0\}.
\end{equation}
\end{teoreema}
\begin{proof}
By \remref{rem-deg}, it suffices to show that for every $\mathbf v\in\R^{nd}$ with $\sum_{i=1}^n v_i=0$
we have
$-\sum_{1\le i < j \le n} \langle v_i, d(x_i-x_j)v_j\rangle>0$ whenever
$x_i\not=x_j$ for all $1\le i<j \le n$.

We have
\begin{equation}
\begin{split}
-\sum_{1\le i < j \le n} \langle &v_i, d(x_i-x_j)v_j\rangle
=-\frac 12\sum_{1\le i,j\le n}\langle v_i, d(x_i-x_j)v_j\rangle \\
&=-\frac C2\int_{\R^d}\re(\sum_{1\le i,j\le n}\frac{1-e^{ik\cdot (x_i-x_j)}}{|k|^{d+\xi}}\langle v_i,(\one-k\otimes k)v_j\rangle)\,dk\\
&=\frac C2\int_{\R^d}\re\langle\sum_{i=1}^n v_i e^{ik\cdot x_i},
\frac{\one-k\otimes k}{|k|^{d+\xi}}\sum_{i=1}^n v_i e^{ik\cdot x_i}
\rangle\,dk\\
\end{split}
\end{equation}

The rest goes as in Proposition 1 of \cite{EX}:
For the integral to be zero, we have to have
\begin{equation}
\sum_{i=1}^nv_i e^{ik\cdot x_i}=\alpha(k)k
\end{equation}
almost everywhere for some scalar function $\alpha$. Taking the
exterior product
(i.e. the antisymmetric part of the tensor product) with respect to
$k$ and Fourier transforming in the sense of distributions we arrive
at
\begin{equation}
\sum_{i=1}^n v_i\wedge\nabla\delta(x-x_n)=0.
\end{equation}
Thus for any smooth test function $\phi$
\begin{equation}
\sum_{i=1}^n v_i\wedge\nabla\phi(x_n)=0.
\end{equation}
This is a contradiction since the values of $\nabla\phi$ can be
arbitrarily specified on a discrete set and the $x_n$'s are all distinct.
\ifcmp\qed\fi
\end{proof}

\subsection{Estimates for the symbol of $M_n$}\label{sec-bnd}
We shall now show that the symbol of $M_n$ can be estimated using
the symbols of $M_m$, $m\in\{2,...,n-1\}$.

\begin{definitio}\label{def-rank}
Let $\mathbf{x}\in \R^{(n-1)d}$. The dimension of the zero eigenspace
of $\sigma(M_n)$ at $\mathbf{x}$ divided by $d$ is called the \emph{rank}
of the point $\mathbf{x}$ and denoted $\rk(\mathbf{x})$. In particular
$\mathbf x$ is a degeneration point of $\sigma(M_n)$ iff $\rk(\mathbf x)>0$. 
\end{definitio}
Below, for a symbol $A$ and invertible linear transformation $L$
we define the symbol $A^L$ by the formula $A^L(x):=LA(L^{-1}x)L^T$.

\begin{teoreema}\label{thm-stru}
Let $n\ge 2$ and $\mathbf x\in\Sph^{(n-1)d-1}$. Then either
$M_n$ is uniformly elliptic in some neighbourhood of $\mathbf x$
or there is a invertible linear transformation $L$ of $\R^{(n-1)d}$, a
neighbourhood $U$ of $L\mathbf x$ so that
$\sigma(M_n)^L\sim \bigoplus_{i=1}^k \sigma(M_{n_i})\oplus\one$
on $U$ with $k\ge 1$, each $n_k\ge 2$, $\rk(\mathbf x)=\sum_{i=1}^k(n_k-1)<n-1$
and $(L\mathbf x)_i=0$ for $1\le i \le\sum_{j=1}^k(n_k-1)$.
\end{teoreema}

Let's introduce some convenient notation at this point. First
of all $[i,j]:=\{i,...,j\}$. Let $A\subseteq[1,n]$. Then we write
\begin{equation}
\begin{split}
&x_A:=\sum_{i\in A}x_i\\
&\gamma_A:=\langle v_{\min A},(d(x_A)-d(x_{A\setminus\{\min A\}})\\
&\qquad\qquad -d(x_{A\setminus\{\max A\}})+d(x_{A\setminus\{\min A,\max A\}}))v_{\max A}\rangle\\
&\sigma_A:=\sum_{i,j\in A; i\le j}\gamma_{A\cap[i,j]}.
\end{split}
\end{equation}

Moreover $x_{i,j}:=x_{[i,j]}$, $\sigma_{i,j}:=\sigma_{[i,j]}$,
$\gamma_{i,j}:=\gamma_{[i,j]}$ and $\sigma_{i}:=\gamma_{i}:=\gamma_{\{i\}}$.

\subsection{Two propositions for the proof of \thmref{thm-stru}}\label{subsec-lem}

Our purpose here is to prove \proref{pro-don} and \proref{pro-messest}.
Let us illustrate what we're going to do by studying $\sigma(M_3)$ in some
detail. 

Let $\mathbf x\in\Sph^{2d-1}$ be such that $x_1=0$, i.e. $\mathbf x=(x_1,x_2)$
with $x_2\in\Sph^{d-1}$. We'll show that there is a neighbourhood $U$ of
$\mathbf x$ and $C<\infty$ so that for every $\mathbf y\in U$ we have
\begin{equation}\label{equ-m3}
\frac 1C\bigl(|y_1|^\xi|v_1|^2+|y_2|^\xi|v_2|^2\bigr)\le
\sigma(M_3)(\mathbf y)\le C\bigl(|y_1|^\xi|v_1|^2+|y_2|^\xi|v_2|^2\bigr).
\end{equation}

Let $E$ be given by \lemref{lem-cro1} and let $\epsilon\in(0,\frac 14)$ be
such that
\begin{equation}
E\bigl((2\epsilon)^{1-\xi/2}+(2\epsilon)^{\xi/2}\bigl)\le \frac 12
\end{equation}
and let
\begin{equation}
U:=B(0,\epsilon)\times \{\frac 12< |y_2|< \frac 32\}.
\end{equation}

By our choice of $\epsilon$ we have
\begin{equation}
|\gamma_{1,2}|\le\frac 12(|y_1|^\xi|v_1|^2+|y_2|^\xi|v_2|^2)
\end{equation}
in $U$. In other words \equref{equ-m3} holds and thus
$\sigma(M_3)\sim\sigma(M_2)\oplus\one$ on $U$.

\proref{pro-don} will be used when we have several (or all) coordinates
away from the degeneration set. As might be guessed from our
calculation with $\sigma(M_3)$, the point of
Lemmata \ref{lem-cro1}-\ref{lem-cro4} is that in the proof
of \thmref{thm-stru} we need to have estimates for the
crossterms with the flavor 
\begin{equation}
|\gamma_{i,j}|\le\text{ something }\cdot(|x_i|^\xi|v_i|^2+|x_j|^\xi|v_j|^2).
\end{equation}
We have neatly blackboxed all this mess into \proref{pro-messest}; the
Lemmata of this section are not directly used in the proof of
\thmref{thm-stru}. The proofs can be found in \appref{app-lemm}.

\begin{propositio}\label{pro-don}
Suppose $n\ge 1$, $\epsilon\in(0,1)$ and
let
\begin{equation}
A:=\{\mathbf x\in\R^{nd}:\epsilon\max\{|x_{i,j}|:1\le i\le j\le n\}
\le\min\{|x_{i,j}|:1\le i\le j\le n\}\}.
\end{equation}
Then there is $C<\infty$ so that for every $\mathbf x\in A$ we have
\begin{equation}
\frac 1C\sum_{i=1}^n |x_i|^\xi|v_i|^2
\le \sigma(M_{n+1})
\le C\sum_{i=1}^n |x_i|^\xi|v_i|^2
\end{equation}
\end{propositio}

\begin{apulause}\label{lem-cro1}
There is a constant $E<\infty$ such that if $1\le i<n$ and
$|x_i|<\frac{1}{2}|x_{i+1}|$, then
\begin{equation}
\begin{split}
|\langle v_i,&(d(x_i+x_{i+1})-d(x_i)-d(x_{i+1}))v_{i+1}\rangle|\\
&\le E\biggl(\bigl(\frac{|x_i|}{|x_{i+1}|}\bigr)^{1-\xi/2}+
\bigl(\frac{|x_i|}{|x_{i+1}|}\bigr)^{\xi/2}\biggr)(|x_i|^\xi|v_i|^2+|x_{i+1}|^\xi|v_{i+1}|^2).
\end{split}
\end{equation}
\end{apulause}

\begin{apulause}\label{lem-cro2}
There is a constant $E<\infty$ such that if $1\le i<i+1<j\le n$,
$|x_i|<\frac 12\min\{|x_{i+1,j}|,|x_{i+1,j-1}|\}$ and $|x_j|>0$, then
\begin{equation}
\begin{split}
|\langle v_i,&(d(x_{i,j})-d(x_{i+1,j})-d(x_{i,j-1})+d(x_{i+1,j-1}))v_j\rangle|\\
&\le E\biggl(\bigl(\frac{|x_i|}{|x_{i+1,j}|}\bigr)^{1-\xi/2}\bigl(\frac{|x_{i+1,j}|}{|x_j|}\bigr)^{\xi/2}+
\bigl(\frac{|x_i|}{|x_{i+1,j-1}|}\bigr)^{1-\xi/2}\bigl(\frac{|x_{i+1,j-1}|}{|x_j|}\bigr)^{\xi/2}\biggr)\\&\cdot(|x_i|^\xi|v_i|^2+|x_j|^\xi|v_j|^2).
\end{split}
\end{equation}
\end{apulause}

\begin{apulause}\label{lem-cro3}
There is a constant $E<\infty$ such that if $1\le i<i+1<j\le n$,
$\frac 12|x_{i+1,j-1}|\le|x_i|<\frac 12|x_{i+1,j}|$ and $|x_j|>0$, then
\begin{equation}
\begin{split}
|\langle v_i,&(d(x_{i,j})-d(x_{i+1,j})-d(x_{i,j-1})+d(x_{i+1,j-1}))v_j\rangle|\\
&\le E\biggl(\bigl(\frac{|x_i|}{|x_{i+1,j}|}\bigr)^{1-\xi/2}\bigl(\frac{|x_{i+1,j}|}{|x_j|}\bigr)^{\xi/2}+\bigl(\frac{|x_i|}{|x_j|}\bigr)^{\xi/2}\biggr)(|x_i|^\xi|v_i|^2+|x_j|^\xi|v_j|^2).
\end{split}
\end{equation}
\end{apulause}

\begin{apulause}\label{lem-cro4}
There is $E<\infty$ so that if $1\le i<i+1<j\le n$ and
$\max\{|x_i|,|x_j|\}<\frac 13\{|x_{i+1,j-1}|\}$, we have
\begin{equation}
\begin{split}
|\langle v_i,&(d(x_{i,j})-d(x_{i+1,j})-d(x_{i,j-1})+d(x_{i+1,j-1}))v_j\rangle|\\
&\le E\bigl(\frac{|x_i|}{|x_{i+1,j-1}|}\bigr)^{1-\xi/2}
\bigl(\frac{|x_i|}{|x_{i+1,j-1}|}\bigr)^{1-\xi/2}(|x_i|^\xi|v_i|^2+|x_j|^\xi|v_j|^2).
\end{split}
\end{equation}
\end{apulause}

We still have one more Lemma to go before we can start proving
\proref{pro-messest}. We'll illustrate it with $\sigma(M_6)$.
Let $\mathbf x\in\Sph^{5d-1}$ with $|x_1|=|x_3|=|x_5|=0$ and
$|x_2|,|x_4|,|x_{2,4}|>0$. By \proref{pro-don} $\sigma_{\{2,4\}}(y_2,y_4)$
behaves like $|y_2|^{\xi}|v_2|^2+|y_4|^\xi|v_4|^2$ in a neighbourhood
of $(x_2,x_4)$. Unfortunately
the relevant part of $\sigma(M_6)$ is $\gamma_2+\gamma_4+\gamma_{2,4}$,
but at least we would have some hope, if we could get an estimate of the form
\begin{equation}
|\gamma_{2,4}-\gamma_{\{2,4\}}|\le\text{ something }\cdot(|y_2|^\xi|v_2|^2+|y_4|^\xi|v_4|^2)
\end{equation}
for $\mathbf y$ in a neighbourhood of $\mathbf x$.

This is the point of \lemref{lem-cro5}. More precisely, let
\begin{equation}
\mu:=\min\{|y_2|,|y_4|,|y_{2,4}|\}\le\max\{|y_2|,|y_4|,|y_{2,4}|\}=:\nu
\end{equation}
and let $C<\infty$ be such that if $\frac \mu 2<|y_2|,|y_4|,|y_2+y_4|<2\nu$
we have
\begin{equation}\label{equ-estm6}
\frac 1C(|y_2|^{\xi}|v_2|^2+|y_4|^\xi|v_4|^2)\le \sigma_{\{2,4\}}\le
C(|y_2|^{\xi}|v_2|^2+|y_4|^\xi|v_4|^2).
\end{equation}

Let $\epsilon\in(0,\frac \mu 6)$ be such that
\begin{equation}
E\biggl(\bigl(\frac{2\epsilon}\mu\bigr)^{1-\xi/2}\bigl(\frac{4\nu}\mu\bigr)^{\xi/2}+\bigl(\frac{2\epsilon}\mu\bigr)^{\xi/2}\biggr)\le\frac 1{2C}.
\end{equation}

Let
\begin{equation}
U:=\{|y_3|<\epsilon\text{ and }\frac \mu2<|y_2|,|y_4|,|y_{2,4}|,|y_2+y_4|<2\nu\}.
\end{equation}
By \lemref{lem-cro5} for $\mathbf y\in U$ we have
\begin{equation}\label{equ-xcacro5}
|\gamma_{2,4}-\gamma_{\{2,4\}}|\le \frac 1{2C}(|y_2|^{\xi}|v_2|^2+|y_4|^\xi|v_4|^2).
\end{equation}
Combining \equref{equ-xcacro5} with \equref{equ-estm6} we
conclude that $\gamma_2+\gamma_4+\gamma_{2,4}$ behaves like
$|y_2|^{\xi}|v_2|^2+|y_4|^\xi|v_4|^2$ in $U$.

Again, the proof of the following Lemma can be found in \appref{app-lemm}.

\begin{apulause}\label{lem-cro5}
There is $E<\infty$ such that if $1\le i<j\le n$ and
$\{i,j\}\subseteq A\subseteq[i,j]$ and if
$\sum_{k\in[i,j]\setminus A}|x_k|\le\frac 12\min\{|x_{k,l}|:k,l\in A, k\le l\}$
Then
\begin{equation}
\begin{split}
|\gamma_{i,j}-\gamma_A|&\le
E\biggl(\bigl(\frac{\sum_{k\in[i,j]\setminus A}|x_k|}{\min\{|x_{k,l}|:k,l\in A, k\le l\}}\bigr)^{1-\xi/2}\bigl(\frac{\sum_{k\in A}|x_k|}{|x_j|}\bigr)^{\xi/2}\\&+\bigl(\frac{\sum_{k\in[i,j]\setminus A}|x_k|}{\min\{|x_{k,l}|:k,l\in A, k\le l\}}\bigr)^{\xi/2}\biggr)(|x_i|^\xi|v_i|^2+
|x_j|^\xi|v_j|^2).
\end{split}
\end{equation}
\end{apulause}

If $L\in\GL(\R^{(n-1)d})$, we shall use the following somewhat weird
notation: If $\mathbf x\in\R^{(n-1)d}$, we let $Lx_i:=(L\mathbf x)_i$
for $1\le i\le n-1$. Similarly, we let $Lx_{i,j}:=(L\mathbf x)_{i,j}$
for $1\le i\le j\le n-1$.

\begin{remarkki}\label{rem-reorg}
Let $\mathbf x$ be a degeneration point of $\sigma(M_n)$. We claim
that there is a symmetry $L\in \mathcal L_n$ and $A\subsetneq\{1,...,n-1\}$
so that $|Lx_i|=0$ if $i\in A$ and $Lx_{i,j}>0$ if $\{i,...,j\}\not\subseteq A$.
This is easy to see, if we look at the original symbol
$\sigma(\mathcal M_n^{sc})$. Then the claim above simply says that
if we have points $y_1,...,y_n\in\R^d$, then there is a permutation
$\pi\in S_n$ so that if $y_{\pi(i)}=y_{\pi(j)}$ with $\pi(i)\le\pi(j)$,
then $y_{\pi(i)}=y_k$ with every $k$ with $\pi(i)\le k\le \pi(j)$.
Still in other words: if we pick $n$ possibly coinciding points
from $\R^d$, we can label them with numbers $1,...,n$ so that
the coinciding points get consecutive numbers as labels.
\end{remarkki}

Given $x$ and $A$ as above, write $A$ as
\begin{equation}
\{i_1,...,j_1\}\cup...\cup\{i_m,...,j_m\}
\end{equation}
with $i_1\le j_1<j_1+1<i_2\le...<i_m\le j_m$ and
write $\sigma(M_n)$ as
\begin{equation}
\sigma(M_{n})=\sum^m_{l=1}\sigma_{i_l,j_l}+\sigma_{A^c}
+\sum_{i,j\in A^c}\gamma_{i,j}-\sigma_{A^c}+\text{ the rest.}
\end{equation}

Let $\mu:=\min\{|x_{i,j}|:\{i,...,j\}\not\subseteq A\}$ and
$\nu:=\max\{|x_{i,j}|:\{i,...,j\}\not\subseteq A\}$.

\begin{propositio}\label{pro-messest}
For any $C>0$ there is a neighbourhood $U$ of $\mathbf x$ so that
\begin{equation}\label{equ-messest}
|\sum_{i,j\in A^c}\gamma_{i,j}-\sigma_{A^c}+\text{ the rest }|\le\frac 1{2C}\sum_{i=1}^n
|y_i|^\xi|v_i|^2
\end{equation}
for any $\mathbf y\in U$.
\end{propositio}

\begin{proof}
For $\epsilon>0$ let
\begin{equation}
U^\epsilon:=\{\mathbf y\in\R^{nd}:|y_{i,j}|<\epsilon\text{ if }\{i,...,j\}\subseteq A
\text{ and }\mu/2<|y_{i,j}|<2\nu\text{ otherwise}\}.
\end{equation}
Let $N:=\frac{n(n-1)}2$ be the number of terms in $\sigma(M_n)$. We'll
find $\epsilon>0$ so that each term in \equref{equ-messest}
is $\le\frac 1{2NC}\sum_{i=1}^n|x_i|^\xi|v_i|^2$ where
we count each $\gamma_{i,j}-\gamma_{A^c\cap[i,j]}$ with $i,j\in A^c$
as one term.

A (long) moment's look at Lemmata \ref{lem-cro1}-\ref{lem-cro5} reveals us that
this is possible. Here's a list of the requirements for $\epsilon$.
\begin{enumerate}
\item \lemref{lem-cro1}: $\epsilon<\frac\mu 4$ and 
$E((\frac {2\epsilon}{\mu})^{1-\xi/2}+(\frac {2\epsilon}{\mu})^{\xi/2})
\le\frac 1{2NC}$
\item \lemref{lem-cro2}: $\epsilon<\frac\mu 4$ and
$2E(\frac{2\epsilon}\mu)^{1-\xi/2}(\frac{4\nu}\mu)^{\xi/2}\le\frac 1{2NC}$.
\item \lemref{lem-cro3}: $\epsilon<\frac\mu 4$ and
$E((\frac{2\epsilon}\mu)^{1-\xi/2}(\frac{4\nu}\mu)^{\xi/2}+
(\frac{2\epsilon}\mu)^{\xi/2})\le\frac 1{2NC}$
\item \lemref{lem-cro4}: $\epsilon<\frac\mu 6$ and
$E(\frac{2\epsilon}\mu)^{2-\xi}\le\frac 1{2NC}$
\item \lemref{lem-cro5}: $n\epsilon<\frac\mu 4$ and $E((\frac{2n\epsilon}\mu)^{1-\xi/2}(\frac{4n\nu}{\mu})^{\xi/2}+(\frac {2n\epsilon}\mu)^{\xi/2})\le\frac 1{2NC}$.
\end{enumerate}
\ifcmp\qed\fi
\end{proof}

\subsection{The proof of \thmref{thm-stru}}\label{sec-proof-stru}

\begin{proof}(of \thmref{thm-stru})
We shall prove this Theorem by induction on $n$ and we shall 
accomplish this by proving in parallel that there is a
constant $C<\infty$ so that for any $\mathbf x\in\R^{(n-1)d}$ there
is $K\in \mathcal L_n$ so that
\begin{equation}
\frac 1C\sum_{i=1}^{n-1}|Kx_i|^\xi|v_i|^2\le \sigma(M_n)
\le C\sum_{i=1}^{n-1}|Kx_i|^\xi|v_i|^2.
\end{equation}

This is trivial for $\sigma(M_2)$. We assume now that the claim above is
true for $\sigma(M_m)$, $2\le m< n$ and prove it for $\sigma(M_n)$.
This is done as follows. For every $\mathbf x\in \Sph^{nd-1}$ we
find a neighbourhood $U_{\mathbf x}$ of $\mathbf x$ so that
the claim above holds on $U_{\mathbf x}$ with a constant
$C(\mathbf x)$ depending on $\mathbf x$ . Since $\Sph^{nd-1}$
is compact, there is  a finite set $\{\mathbf x_1,...,\mathbf x_k\}$
so that $\Sph^{nd-1}\subseteq\bigcup_{i=1}^k U_{\mathbf x_k}$, so
the claim above will then hold with $C=\max_{1\le i\le k}C(\mathbf x_i)$.

If $\mathbf x$ is not a degeneration point of $M_{n+1}$, then
by \proref{pro-don} the estimate above can be satisfied in a neighbourhood
of $\mathbf x$ with $K=\one$, so we assume $\mathbf x$ is a degeneration point.

We now apply the symmetry discussed in \remref{rem-reorg}, so
we can assume there is nonempty $A\subsetneq\{1,...,n\}$ so that
$|x_i|=0$ if $\,i\in A$ and $|x_{i,j}|>0$ if
$\{i,...,j\}\not\subseteq A$. Write $A$ as
$\{i_1,...,j_1\}\cup...\cup\{i_m,...,j_m\}$ with
$i_1\le j_1<j_1+1<i_2<...<i_m\le j_m$. Denote $A^c:=\{1,...,n\}\setminus A$.
We may even assume that $i_1=1$ and if $m>1$, we have $j_m=n$.
Note that $\rk(\mathbf x)=\#(A)$. Let $U'$ be the neighbourhood
of $\mathbf x$ given by \proref{pro-messest}.

Recall that $\mu$ and $\nu$ were defined as $\mu:=\min\{|x_{i,j}|:\{i,...,j\}\not\subseteq A\}$ and
$\nu:=\max\{|x_{i,j}|:\{i,...,j\}\not\subseteq A\}$. Let
\begin{equation}\label{equ-thebee}
U:=U'\cap\{\frac\mu 2< |y_B|< 2\nu:B\not\subseteq A\}.
\end{equation}

First of all, let $C<\infty$ be such that our induction hypothesis
is satisfied with it for $2\le m< n$ and also that $C$ is so large
that the conclusion of \proref{pro-don} holds with
$\epsilon:=\frac{\mu}{4\nu}$. Also we require that
\begin{equation}\label{equ-sig3}
\frac 1C\max_{B\not\subseteq A}|y_B|^\xi\le 1 \le C\min_{B\not\subseteq A}|y_B|^\xi
\end{equation}
holds whenever $\mathbf y\in U$.

We claim that on $U$ we have
$\sigma(M_{n})\sim \sigma(M_{j_1+1})\oplus\one$ if $m=1$ and
$\sigma(M_{n})\sim \sigma(M_{j_1+1})\oplus
1\oplus\sigma(M_{j_2-i_2+2})\oplus ...\oplus\one\oplus \sigma(M_{n-i_m+2})$ otherwise. Denote the right-hand sides
of these expressions collectively as $\Sigma$.

By our induction hypotheses, for any $\mathbf y'\in U$ and any 
$k\in \{1,...,m\}$ there is a symmetry $K\in \mathcal L_n$
so that for $1\le k\le m$ we have
\begin{equation}\label{equ-sig4}
\frac 1C \sum_{i=i_k}^{j_k}|Ky'_i|^\xi|v_i|^2
\le \sigma(M_{j_k-i_k+2})(Ky'_{i_k},...,Ky'_{j_k})\le C\sum_{i=i_k}^{j_k}|Ky'_i|^\xi|v_i|^2
\end{equation}
with $C$ not depending on $\mathbf y'$: Just pick such a symmetry
$K_k\in\mathcal L_{j_k-i_k+2}$ for $k\in\{1,...,m\}$ and take any
$K\in\mathcal L_n$ such that the restriction to the 
$y_{i_k},...,y_{j_k}$ coordinates is $K_k$. Here we have been abusing
notation with the $K_k$'s so that $K_k$ above operates on coordinates
$y_{i_k},...,y_{j_k}$ and not $y_1,...,y_{j_k-i_k+1}$. Extend
$K_k$ now naturally to whole of $\R^{(n-1)d}$. We can now take
$K$ to be say $K=K_1K_2...K_{m-1}K_m$.

Now for every $\mathbf y'\in U$ fix such a transformation $K_{\mathbf y'}$
and denote $\mathbf y:=K_{\mathbf y'}\mathbf y'$. 

By \equref{equ-sig3} and \equref{equ-sig4} we have
\begin{equation}\label{equ-sim1}
\frac 1C \sum_{i=1}^{n-1}|y_i|^\xi|v_i|^2\le \Sigma(\mathbf y)\le C\sum_{i=1}^{n-1}
|y_i|^\xi|v_i|^2.
\end{equation}

As before, we write 
\begin{equation}
\sigma(M_n)=\sum^m_{l=1}\sigma_{i_l,j_l}+\sigma_{A^c}
+\sum_{i,j\in A^c}\gamma_{i,j}-\sigma_{A^c}+\text{ the rest.}
\end{equation}
The first two terms satisfy
\begin{equation}
\frac 1C\sum_{i=1}^{n-1}|y_i|^\xi|v_i|^2\le\sum^m_{l=1}\sigma_{i_l,j_l}+\sigma_{A^c}\le C\sum_{i=1}^{n-1}|y_i|^\xi|v_i|^2,
\end{equation}
and by \proref{pro-messest} we have
\begin{equation}
|\sum_{i,j\in A^c}\gamma_{i,j}-\sigma_{A^c}+\text{ the rest}|\le \frac 1{2C}\sum_{i=1}^n|y_i|^\xi|v_i|^2.
\end{equation}

So we have
\begin{equation}
\frac 1{2C}\sum_{i=1}^{n-1}|y_i|^\xi|v_i|^2\le \sigma(M_n)(\mathbf y)
\le (C+\frac 1{2C})\sum_{i=1}^{n-1}|y_i|^\xi|v_i|^2.
\end{equation}

Let $U^K:=\{\mathbf y'\in U: K_{\mathbf y'}=K\}$. Clearly
$U=\bigcup\{U^K:K\in\mathcal K_n\}$. We just proved that for any
$\mathbf y'\in U$ we have $\sigma(M_n)\sim \Sigma$ in
$K_{\mathbf y'}U^{K_{\mathbf y'}}$. Since both $\Sigma$ and $\sigma(M_n)$
are invariant under $K_{\mathbf y'}^{-1}$ for any $\mathbf y'\in U$,
we can conclude by \remref{rem-sym} that $\sigma(M_n)\sim \Sigma$
on $U^{K_{\mathbf y'}}$. Since $\mathcal L_n$ is finite we can
conclude that $\sigma(M_n)\sim \Sigma$ on $U$.
\ifcmp\qed\fi
\end{proof}

Let
\begin{equation}\label{equ-mcl}
\begin{split}
\mathcal L'_n:=\{L\in \GL(\R^{(n-1)d}):\exists i_1&,j_1,...,i_{n-1},j_{n-1}:\forall x_1,...,x_{n-1}:\\&L((x_1,...,x_{n-1}))=(x_{i_1,j_1},...,x_{i_{n-1},j_{n-1}})\}.
\end{split}
\end{equation}

Obviously, $\mathcal L'_n$ is a finite set. It is also easy to see that it is a group.
Note that the $L$ as constructed
in \thmref{thm-stru} belongs to $\mathcal L'_n$.

\begin{remarkki}
The following Proposition simply says the following:
Suppose we have a symbol of the form
\begin{equation}
\bigoplus_{i=1}^k\sigma(M_{n_i+1})\oplus\one.
\end{equation}
This corresponds to a splitting $\R^{nd}=\R^{ld}\oplus\R^{(n-l)d}$
with $l=n_1+...+n_k$. Then we can replace $\R^{(n-l)d}$ with
any complementary subspace to $\R^{ld}$ and the symbol looks
the same in these new coordinates as looks the symbol in 
an neighbourhood of $0$ which is bounded in the $\R^{ld}$-direction.
\end{remarkki}

\begin{propositio}\label{pro-coord}
Let $\sigma\sim\bigoplus_{i=1}^k\sigma(M_{n_i+1})\oplus\one$ on a set
$U\subseteq B\times \R^{(n-l)d}$ with $B$ bounded and
$l:=\rk(0)=\sum_{i=1}^k n_i$. Let $L\in\GL(\R^{nd})$ be such that
\begin{enumerate}
\item $L:\{0\}\times\R^{(n-l)d}=\{0\}\times\R^{(n-l)d}$ and
\item Let $P:\R^{nd}\rightarrow\R^{ld}$ be the natural projection onto
the first $ld$ coordinates and let $L':=L\upharpoonright\R^{ld}\times\{0\}$.
Then 
\begin{equation}
\bigl(\bigoplus_{i=1}^k\sigma(M_{n_i+1})\bigr)^{L'}\sim\bigoplus_{i=1}^k\sigma(M_{n_i+1}).
\end{equation}
\end{enumerate}

With these assumptions
\begin{equation}\label{equ-simil}
\sigma^L\sim\bigoplus_{i=1}^k\sigma(M_{n_i+1})\oplus\one
\end{equation}
on $LU$.
\end{propositio}

\begin{proof}
Without loss of generality we may assume that
\begin{equation}
L:=
\begin{pmatrix}
\one &0\\
M &\one\\
\end{pmatrix},
\end{equation}
with $M$ an $\R^{(n-l)d}\times\R^{ld}$-matrix.

Also without loss of generality we may assume $U=B(0,1)\times\R^{(n-l)d}$.

Let $A:=\bigoplus_{i=1}^k\sigma(M_{n_i+1})$. Denote $v:=(v_1,v_2)$ and
$x:=(x_1,x_2)$ where $v_1,x_1\in \R^{ld}$ and $v_2,x_2\in\R^{(n-l)d}$.
Then 
\begin{equation}
\begin{split}
\langle v,(A\oplus\one)^L(x) v\rangle&=\langle v_1,A((L^{-1}x)_1)v_1\rangle
+\langle v_1,A((L^{-1}x)_1)M^Tv_2\rangle+\\
&+\langle M^Tv_2,A((L^{-1}x)_1)v_1\rangle
+|v_2|^2=:(*).
\end{split}
\end{equation}

Since $A(x)$ is a symmetric matrix for every $x$ the two middle terms are
equal. Moreover, $(L^{-1}x)_1=x_1$ and thus
\begin{equation}
(*)=\langle v_1,A(x_1)v_1\rangle+2\langle v_1,A(x_1)M^Tv_2\rangle+|v_2|^2=:(**)
\end{equation}

Next, we use induction on $\rk 0=n_1+...+n_k$. If $\rk 0=1$, i.e.
$A=\sigma(M_2)$ we have
\begin{equation}\label{equ-ineq}
\frac 1C(|v_1|^2+|v_2|^2)\le (**)\le C(|v_1|^2+|v_2|^2)
\end{equation}
for some $C<\infty$ when $(x_1,x_2)\in \Sph^{d-1}\times\R^{(n-1)d}$.
Adding $(|x_1|^{-\xi}-1)|v_2|^2$ and multiplying by $|x_1|^\xi$ yields
\begin{equation}
\frac 1C(|x_1|^\xi|v_1|^2+|v_2|^2)\le (**)\le C(|x_1|^\xi|v_1|^2+|v_2|^2)
\end{equation}
when $(x_1,x_2)\in B(0,1)\times\R^{(n-1)d}$. Since $\sigma(M_2)\sim|\cdot|^\xi$
we can conclude our claim.

Next, suppose our Proposition is true for configurations of rank $<l$
and we prove our claim when $\rk 0=l$. Now cover $\Sph^{ld-1}$ by
finitely many open sets $B_1,...,B_m$ so that
\begin{equation}
\bigl(\bigoplus_{i=1}^{k}\sigma(M_{n_i+1})\bigr)^{L_j}\sim\bigoplus_{i=1}^{k_j}
\sigma(M_{n_{j,i}+1})\oplus\one
\end{equation}
on $B_j$ with some linear transformation $L_j$ and with $\sum_{i=1}^{k_j}n_{j,i}<l$.

Letting $L'_j:=L(L_j\oplus\one)$, and applying this Theorem
on $U_j:=B_j\times\R^{(n-l)d}$ we see that
\begin{equation}
\sigma(M_{n+1})^{L'_j}\sim\bigoplus_{i=1}^{k_j}\sigma(M_{n_{j,i}+1})\oplus\one
\end{equation}
on $U_j$.

Now a similar argument as above for rank $0$ yields the desired
conclusion. The reader may fill in the details.
\end{proof}

The following is an immediate corollary to this proposition.

\begin{korollaari}\label{cor-coord}
Let $L\in \mathcal L$ be such that for some neighbourhood $U$ of $x$ we
have
\begin{equation}
\sigma(M_{n+1})^L\sim\bigoplus_{i=1}^k\sigma(M_{n_i+1})\oplus\one
\end{equation}
on $LU$. Then for every $L'\in \mathcal L$ such that 
\begin{equation}
L^{-1}=L^{\prime-1}\text{ on }\{|x_i|=0:1\le i\le\rk x\}
\end{equation}
we have
\begin{equation}
\sigma(M_{n+1})^{L'}\sim\bigoplus_{i=1}^k\sigma(M_{n_i+1})\oplus\one
\end{equation}
on $L'U$.
\end{korollaari}

\subsection{Some Corollaries}

\begin{korollaari}\label{cor-low}
For every $n\ge 2$ there is $C>0$ such that
\begin{equation}
C d(\mathbf x,\Dgn(M_{n}))^\xi\le \sigma(M_{n}).
\end{equation}
\end{korollaari}

The proof of this fact is easy and thus omitted. The assumptions
of \thmref{thm-har} are now satisfied (by \corref{cor-low},
\thmref{thm-poin} and \proref{pro-poin}) for $M_n$. Moreover,
we can directly calculate the dimension of $M_n$:

\begin{korollaari}\label{cor-dim}
There is $C<\infty$ such that for any $f\in L^2(\R^{(n-1)d})$ we have
\begin{equation}\label{equ-dim}
||e^{-M_nt}f||_\infty\le C t^{-\frac{(n-1)d}{4-2\xi}}||f||_2.
\end{equation}
Moreover, $C$ depends only on the lower bound for $\sigma(M_n)$.
\end{korollaari}

\begin{proof}
By \proref{pro-poin} there is $C<\infty$ so that
\begin{equation}
||f||_q\le C||d(\mathbf x,\Dgn(M_{n}))^{\xi/2}\nabla f||_2=:(*)
\end{equation}
for any $f\in C^\infty_0(\R^{(n-1)d})$ with $q:=\frac{2n}{n+\xi-2}$.

By \corref{cor-low} we have
\begin{equation}
(*)\le C'\langle f,M_n f\rangle.
\end{equation}

Finally, by \thmref{thm-sou} we can conclude that \equref{equ-dim}
holds. 
\ifcmp\qed\fi
\end{proof}

\begin{korollaari}\label{cor-glob}
For any $\rho\in(0,1)$ there is $C<\infty$ such that for any
$\mathbf x\in\R^{(n-1)d}$ and any $\mathbf y\not\in B(\mathbf x,\rho|\mathbf x|)$ we have
\begin{equation}
K_{M_n}(t,\mathbf x,\mathbf y)\le Ct^{-\frac{(n-1)d}{2-\xi}}\exp\{-\frac{|\mathbf x-\mathbf y|^{2-\xi}}{Ct}\}
\end{equation}
and
\begin{equation}
G_{M_n}(\mathbf x,\mathbf y)\le C|\mathbf x-\mathbf y|^{2-\xi-(n-1)d}.
\end{equation}
\end{korollaari}

\begin{proof}
This is a direct consequence of \proref{pro-met}, \thmref{thm-gub} and
\corref{cor-dim}.
\ifcmp\qed\fi
\end{proof}

\begin{korollaari}\label{cor-prod}
Suppose $A\sim^\lambda \sigma(M_{n_1+1})\oplus...\oplus\sigma(M_{n_k+1})\oplus\one$
on $\R^{ld}\times\R^{(n-l)d}$ with $l:=n_1+...+n_k<n$ and
let $\epsilon>0$ be given. Then there is $C<\infty$
such that if $z\not\in B(y_1,\epsilon|y_1|)\times B(y_2,\epsilon|y_1|^{1-\xi/2})$ (here $y:=(y_1,y_2)\in\R^{ld}\times\R^{(n-l)d}$), we have
\begin{equation}
K_A(t,y,z)\le C t^{-\frac{ld}{2-\xi}-\frac{n-l}{2}}
\exp\{-\frac{|y_1-z_1|^{2-\xi}+|y_2-z_2|^2}{Ct}\}.
\end{equation}
Moreover $C$ depends on $A$ only through $\lambda$, $n_1,...,n_k$ and $n$.
\end{korollaari}

\begin{proof}
The proof is straightforward using \thmref{thm-prod},
\proref{pro-met} and \corref{cor-dim} and we leave the details
for the reader. The only finesse is the
appearance of $B(y_2,\epsilon|y_1|^{1-\xi/2})$ above. This
is due to the fact that if $z_1\in B(y_1,\epsilon|y_1|)$
and $z_2\not\in B(y_2,\epsilon|y_1|^{1-\xi/2})$, we have
\begin{equation}
\begin{split}
|y_1-z_1|^{2-\xi}+|y_2-z_2|^2&\le (\epsilon|y_1|)^{2-\xi}+|y_2-z_2|^2\\
&\le \epsilon^{-\xi}|y_2-z_2|^2+|y_2-z_2|^2.
\end{split}
\end{equation}
\ifcmp\qed\fi
\end{proof}

\ifthesis\pagestyle{headings}\fi
\section{Local estimates for the heat kernel}\label{sec-local}
The main result in this section is \thmref{thm-loc}. Superficially
it is very similar to \corref{cor-prod}, but there is a very
important difference: In \corref{cor-prod} one assumes that
\begin{equation}\label{equ-sim}
A\sim\sigma(M_{n_1+1})\oplus...\oplus\sigma(M_{n_k+1})\oplus\one
\end{equation}
in $\R^{nd}$ but in \thmref{thm-loc} $A=\sigma(M_{n+1})$
and \equref{equ-sim} holds only in a relatively compact
neighbourhood of a point $x$. The point of this section is to
close the gap between these two results. We start with some
technicalities and prove a uniform version of the Harnack
inequality adapted to our case.

\begin{remarkki}
In a few places we use the somewhat terse assumption
``$A$ has a heat kernel''. In these places we assume that
$A$ has a heat kernel $K$ such that both $K(\cdot,x,\cdot)$ and
$K(\cdot,\cdot,x)$ are solutions to
$u_t+Au=0$ in the sense of \remref{rem-solution} and that for every
$t$ and $x$ we have both
\begin{equation}
\int dy\,K(t,x,y)\le 1\text{ and }\int dy\,K(t,y,x)\le 1.
\end{equation}
In the cases that are of interest to us (see \remref{rem-kernel})
this is the case and moreover our heat
kernels are symmetric in the spatial coordinates.
\end{remarkki}

A well-known argument (see for example \cite{VSC}, section I.3, page 5) yields
the following: Suppose $A$ is a divergence-form operator on $\R^n$ with a 
nonnegative symbol. Suppose also that $A$ is uniformly elliptic on some ball
$B$ and that $A$ has a heat kernel. Then for any ball $B'\subset\subset B$
there is $C<\infty$ such that we have
\begin{equation}
K(t,x,y)\le C t^{-n/2}
\end{equation}
whenever $t\in(0,1]$, $x\in B'$ and $y\in\R^d$.
We shall now make a generalization (\corref{cor-lub1})
of this result.

So for the rest of the section we fix a symbol $A$ on $\R^{nd}$ and
suppose that
\begin{equation}
A\sim^\lambda \sigma(M_{n_1+1})\oplus...\oplus\sigma(M_{n_k+1})\oplus\one
\end{equation}
on $B(0,2)\times B(0,2)\subseteq \R^{ld}\times\R^{(n-l)d}$, where
$l:=n_1+...+n_k$. Let's denote
\begin{equation}
Q:=\overline{B}(0,1)\times \overline{B}(0,1)\text{ and }D:=\Sph^{nd-1}\times \overline{B}(0,1).
\end{equation}

\begin{propositio}\label{pro-har2}
For each $t\in(0,1]$ there is an open covering $\{U^t_y\}_{y\in Q}$
of $Q$ with the following properties:
\begin{enumerate}
\item \label{item-hartri} $y\in U^t_y$ for every $y\in Q$ and $t\in(0,1]$.
\item \label{item-harcont} There is $\epsilon>0$ not depending on $t$ such that $B(y_1,\epsilon t^{1/2-\xi})\times
B(y_2,\epsilon \sqrt{t})\subseteq U^t_y$
\item \label{item-harhar} For every $t\in(0,1]$, every $y\in Q$ and
every positive solution
$u$ of \linebreak $u_t=\nabla\cdot A\nabla u$ on $(0,3)\times U^t_y$ we have 
\begin{equation}
\sup_{y'\in U^t_y}u(t,y')\le C\inf_{y'\in U^t_y}u(2t,y').
\end{equation}
Moreover, $C$ depends on $A$ only through $\lambda$, $n_1,...,n_k$ and $n$.
\end{enumerate}
\end{propositio}

\begin{remarkki}
Strictly speaking in \itemref{item-harhar} we only assume $u$ is a solution
of \linebreak $u_t=\nabla\cdot A\nabla u$ in the sense of \remref{rem-solution}
on $(\epsilon,3)\times U^t_y$ for every $\epsilon\in(0,3)$.
\end{remarkki}

\begin{korollaari}\label{cor-har2}
\proref{pro-har2} holds with obvious modifications for any affine
transform $A^K$ of $A$ with possibly different $\epsilon$ and $C$.
\end{korollaari}

To give some intuition to the reader we first give a Corollary to
this Proposition.

\begin{korollaari}\label{cor-lub1}
There is $C<\infty$ such that
\begin{equation}
K_{A}(t,y,y')\le C t^{-\frac{ld}{2-\xi}-\frac{(n-l)d}{2}}
\end{equation}
for any $y\in Q$, $y'\in \R^{nd}$ and $t\in(0,1]$.
\end{korollaari}

\begin{proof}
By \proref{pro-har2} for any $y\in Q$ and
$y'\in \R^{nd}$ we have 
\begin{equation}
\begin{split}
t^{\frac{ld}{2-\xi}+\frac{(n-l)d}{2}}K_A(t,y,y')
&\le C'|U^t_y|\sup_{y''\in U^t_y}K_A(t,y'',y')\\
&\le CC'|U^t_y|\inf_{y''\in U^t_y}K_A(2t,y'',y')\\
&\le CC'\int_{U^t_y}K_A(2t,y'',y')\,dy''\\
&\le CC'.
\end{split}
\end{equation}
\ifcmp\qed\fi
\end{proof}

Next we prove a small Lemma used in the proof of \proref{pro-har2}.
The setup here is the following. Let
$y\in D$. In our proof of \proref{pro-har2} we use induction on
rank. By \thmref{thm-stru} there is an invertible affine transformation
$K_y$ of $\R^{nd}$ sending $y$ to $0$ so that
\begin{equation}
A^{K_y}\sim \sigma(M_{n'_1+1})\oplus ...\oplus\sigma(M_{n'_k+1})\oplus\one
\end{equation}
on $B(0,2)\times B(0,2)$ with $l':=n'_1+...+n'_k<l$. Now 
\lemref{lem-smalltech} allows us to conclude that if
\itemref{item-harcont} of \proref{pro-har2} holds for the covering
associated with $y$ in $K_y$-coordinates with some $\epsilon$
(for convenience, we have put this $\epsilon$ equal to $1$ in the
statement of \lemref{lem-smalltech}), then it holds
in the usual coordinates of $\R^{nd}$ with some other $\epsilon$.

Here is our choice of the subspaces for \lemref{lem-smalltech}:
\begin{enumerate}
\item $S_1:=K_y^{-1}[\R^{l'd}\times\{0\}]-\{y\}$ and
\item $S_2:=K_y^{-1}[\{0\}\times\R^{(n-l')d}]-\{y\}$.
\end{enumerate}
In other words $S_2$ is the degeneration subspace associated
with $y$. The fact that $y\in Q$ guarantees that
$\{0\}\times \R^{(n-l)d}\subseteq S_2$. Note that the
$-\{y\}$ in the definition of $S_2$ is redundant, since $y\in S_2$,
but we didn't want to confuse the reader a few lines ago, did we?

\begin{apulause}\label{lem-smalltech}
Let $S_1,S_2$ be a splitting of $\R^{nd}$ into
complementary subspaces so that $\{0\}\times\R^{(n-l)d}\subseteq S_2$.
Assume also that each of them is equipped with a norm and
denote the balls with respect to these norms
with $B_i(x,r)$ with $i=1,2$. Then there
is $\epsilon>0$ so that
\begin{equation}
B(0,\epsilon t^{1/(2-\xi)})\times B(0,\epsilon \sqrt{t})
\subseteq B_1(0,t^{1/(2-\xi)})\times B_2(0,\sqrt{t})
\end{equation}
for any $t\in(0,1]$.
\end{apulause}

\begin{proof}
Obviously there is $\epsilon>0$ so that
\begin{equation}
B(0,\epsilon)\times B(0,\epsilon)
\subseteq B_1(0,1)\times B_2(0,1)
\end{equation}

Let us write $B(0,\epsilon t^{1/(2-\xi)})\times B(0,\epsilon\sqrt{t})$
as
\begin{equation}
B(0,\epsilon t^{\frac{1}{2-\xi}})\times \R^{(n-l)d}\cap B(0,\epsilon \sqrt{t})\times B(0,\epsilon \sqrt{t})
\end{equation}
and similarly for $B_1(0,t^{1/(2-\xi)})\times B_2(0,\sqrt{t})$
(we used the fact that $t^{1/(2-\xi)}\le \sqrt t$ for $t\in (0,1]$).

Now since $\{0\}\times\R^{(n-l)d}\subseteq S_2$, we conclude by scaling that
\begin{equation}
B(0,\epsilon t^{\frac{1}{2-\xi}})\times \R^{(n-l)d}\subseteq B_1(0,t^{\frac 1{2-\xi}})\times S_2.
\end{equation}
for any $t>0$.

Also by scaling we get
\begin{equation}
B(0,\epsilon\sqrt{t})\times B(0,\epsilon\sqrt{t})\subseteq
B_1(0,\sqrt{t})\times B_2(0,\sqrt{t}).
\end{equation}
for any $t>0$.
\ifcmp\qed\fi
\end{proof}

\begin{proof}(of \proref{pro-har2})

If $l=0$, then we just choose
$U^t_y:=B(y,\sqrt t)$. Obviously, these
sets satisfy \itemref{item-harcont} above and by classical results (see again
\cite{VSC}, section I.3, page 5) they satisfy \itemref{item-harhar} too.

Next we assume that the cases $<l$ have been handled and prove the 
Proposition for $l$. This is done in three phases:
\begin{enumerate}
\item Phase 1: Use our induction hypothesis (i.e. that the cases $<l$ have
been handled) to handle points in $D$.
\item Phase 2: Use scaling to handle points $z\in Q$ with
$0<|z_1|<1$ and times $t\in(0,|z_1|^{2-\xi}]$. And finally
\item Phase 3: Do something creative for points $z\in Q$
and times $t\in(|z_1|^{2-\xi},1]$. Note that this includes defining
the sets $U^t_z$ when $|z_1|=0$.
\end{enumerate}

First, phase 1: By compactness, there is $\{y_1,...,y_k\}\subseteq D$
so that \ifcmp{\linebreak[4]}\fi $\{K^{-1}_{y_i}[B(0,1)\times B(0,1)]\}^k_{i=1}$ cover $D$. Obviously
each $y_i$ is of rank $<l$. For each $t\in(0,1]$ and $z\in D$ pick
$U^t_z$ to be one of the $U^t_z$'s associated with some of the
$y_1,...,y_k$ (this is possible by induction hypothesis and \corref{cor-har2}).
Now these $U^t_z$'s satisfy \itemref{item-harcont} and \itemref{item-harhar},
where \itemref{item-harhar} satisfied by induction and \itemref{item-harcont}
is satisfied by \lemref{lem-smalltech} (and the discussion before
it) and finiteness of the set $\{y_1,...,y_k\}$. 

Next, phase 2: We define the sets $U^t_z$ for $z$'s with $0<|z_1|<1$ and
$t\in(0,|z_1|^{2-\xi}]$. This is achieved by scaling
$A$ outwards so that in this scaling $z$ travels to $D$.
Then the symbol $A^z$ obtained this way has the same upper and lower
bounds as $A$ on $B(0,2)\times B(0,2)$, so we can use
our sets $U^t_y$ defined above for $y\in D$. After this we just scale
things back.

So, let $z\in Q$ with $0<|z_1|<1$ and let
\begin{equation}
y^z:=(y_1/|z_1|,z_2+(y_2-z_2)/|z_1|^{1-\xi/2}).
\end{equation}
Let $A^z$ be defined by 
\begin{equation}
A^z_{ij}(y):=
\begin{cases}
|z_1|^{\xi}A_{ij}(y^z)&\text{if $1\le i, j\le ld$}\\
|z_1|^{\xi/2}A_{ij}(y^z)&\text{if $1\le i\le ld < j\le nd$
or }\\&\text{$1\le j\le ld< i\le nd$}\\
\sigma(A_{ij}(y^z)&\text{if $ld< i,j\le nd$}
\end{cases}
\end{equation}

Similarly define $u^z$ by $u^z(t,y):=u(|z_1|^{\xi-2}t,y^z)$. Now
if $u$ satisfies $u_t=\nabla A\cdot\nabla u$ on
$(0,3)\times B(0,2)\times B(0,2)$, then
$u^z$ satisfies $u^z_t=\nabla\cdot A^z\nabla u^z$ on
this same set. Since now if $A\sim^\lambda \sigma(M_{n_1+1})\oplus...\oplus \sigma(M_{n_k+1})\oplus\one$ on $B(0,2)\times B(0,2)$,
then the same is true of $A^z$ we
can conclude that \itemref{item-harcont} and \itemref{item-harhar} hold for $A^z$ with the same constants
as for $A$. So if we scale back and let
\begin{equation}
U^t_z=\{(|z_1|y_1,z_2+|z_1|^{1-\xi/2}(y_2-z_2)):(y_1,y_2)\in U^{|z_1|^{\xi-2}t}_{\hat z}\}
\end{equation}
then \itemref{item-harcont} and \itemref{item-harhar} hold for these
whenever defined.

Finally, phase 3: To finish the argument,
we set for $t\ge |z_1|^{2-\xi}$
\begin{equation}
U^t_z=B(0,\frac 32 t^{1/(2-\xi)})\times B(z_2,\frac 12 \sqrt{t}).
\end{equation}

Now \itemref{item-harcont} holds for these sets. To prove \itemref{item-harhar}
we may assume without
loss of generality that $z_2=0$ and let $A^t$ be defined as follows:
\begin{equation}
A^t_{ij}(y_1,y_2):=
\begin{cases}
t^{-\frac{\xi}{2-\xi}}A_{ij}(y_1 t^{1/(2-\xi)},y_2\sqrt{t})&\text{if $1\le i, j\le ld$}\\
t^{-\frac{\xi}{4-2\xi}}A_{ij}(y_1 t^{1/(2-\xi)},y_2\sqrt{t})&\text{if $1\le i\le ld < j\le nd$
or }\\&\text{$1\le j\le ld< i\le nd$}\\
A_{ij}(y_1 t^{1/(2-\xi)},y_2\sqrt{t})&\text{if $ld< i,j\le nd$}\\
\end{cases}
\end{equation}

As before, for $t\in(0,1]$ the substitution $A\mapsto A^t$
preserves the constant in the Harnack inequality (\thmref{thm-har}) and thus
we can conclude that \itemref{item-harhar} holds.
\ifcmp\qed\fi
\end{proof}

\begin{remarkki}\label{rem-mod}
It is not hard to modify the previous proof so that for given
$\epsilon'>0$ there is $\epsilon>0$ so that
\begin{enumerate}
\item $B(y_1,\epsilon t^{1/(2-\xi)})\times B(y_2,\epsilon
\sqrt{t})\subseteq U^t_y$ for every $t\in (0,1]$ and
\item $U^t_y\subseteq B(y_1,\epsilon't^{1/(2-\xi)})
\times B(y_2,\epsilon'\sqrt{t})$, when $|y_1|^{2-\xi}\le
t\le 1$. 
\item $U^t_y\subseteq B(y_1,\epsilon'|y_1|)\times B(y_2,\epsilon'|y_1|^{(2-\xi)/2})$, when $0<t\le|y_1|^{2-\xi}$.
\end{enumerate}
We need (2) and (3) in the proof of \thmref{thm-loc}. There we need to
find $\epsilon'>0$ so that $U_z^t$ and
$B(y_1,\epsilon't^{1/(2-\xi)})\times B(y_2,\epsilon'\sqrt{t})$ are
disjoint whenever $z\not\in B(y_1,t^{1/(2-\xi)})\times B(y_2,\sqrt{t})$
and this is hard to arrange if we don't have any kind of control over
the $U_z^t$'s from outside. This required control is provided by (2) and
(3) above.  The actual choice of $\epsilon'>0$ is done in \lemref{lem-tech}. 

Anyway, it is quite easy to make (2) and (3) hold. First of all, it
is easy to see that (2) and (3) hold with some $\epsilon'_0>0$
when $U^t_y$'s are defined as in the proof of
\proref{pro-har2}. By letting $V^t_y:=U^{t/T}_y$ with
$T:=(\epsilon'_0/\epsilon')^2$ we see that $V^t_y$'s for $t\in(0,1]$
satisfy (1)-(3) above together with the claims of \proref{pro-har2}. 
The details are left to the reader. We will use \proref{pro-har2} in
this form in the proofs below.
\end{remarkki}

We now have to estimate the tails of the heat kernel. We use
a common probabilistic argument for this (killing probabilities).
Denote
\begin{equation}\label{equ-deedef}
d(x,y)^2:=\max\{|x_1-y_1|^{2-\xi},|x_2-y_2|^2\}.
\end{equation}
Obviously there is $C<\infty$ so that
\begin{equation}
C^{-1}d(x,y)\le\sqrt{|x_1-y_1|^{2-\xi}+|x_2-y_2|^2}\le Cd(x,y)
\end{equation}

Below, $P^y_A(\sup_{s\le t}d(X_s,y)\ge\mu)$
denotes the probability of the diffusion $X$ associated with
$A$ starting from $y$ at time $0$ hitting the set
$\{z:d(y,z)=\mu\}$ before time $t$.

The following is Proposition 6.5 on page 179 of \cite{B}.

\begin{propositio}\label{pro-unik}
Suppose $A\sim^\lambda \one$ on $\R^l$. There is $C<\infty$ depending on
$A$ only through $\lambda$ such that
\begin{equation}
\prob^y_A(\sup_{s\le t}|X_s-y|\ge\mu)\le C\exp\{-\frac{\mu^2}{Ct}\}.
\end{equation}
\end{propositio}

\begin{korollaari}\label{cor-unik}
Suppose $A\sim^\lambda\one$ on $B(0,2)\subseteq \R^{nd}$. Then there is $C<\infty$ depending
on $A$ only through $\lambda$ such that for
every $y\in B(0,1)$, $z\in B(y,\frac 12)$ and $0<t\le 1$ we have
\begin{equation}\label{equ-unik}
K_A(t,y,z)\le C t^{-\frac{nd}{2}}\exp\{-\frac{|y-z|^2}{Ct}\}
\end{equation}
\end{korollaari}

The proof of this Corollary is quite simple and well-known (folklore)
and we shall not prove it here, but the interested reader can reconstruct
the argument from the proof of \thmref{thm-loc} which is a generalization
of \corref{cor-unik}.

Unfortunately we need the following technicality in the proofs of
\proref{pro-kil} and \thmref{thm-loc}.

\begin{apulause}\label{lem-tech}
Suppose $\epsilon''>0$ is given. Then there is $\epsilon'>0$ so that
if $d(y,z)\ge\epsilon''|y_1|^{1-\xi/2}$, we have
\begin{equation}\label{equ-tech1}
\{z':d(z,z')\le\epsilon'|z_1|^{1-\xi/2}\}\subseteq\{z':d(z,z')\le \frac{d(y,z)}2\}
\end{equation}
and
\begin{equation}\label{equ-tech2}
B(y_1,\epsilon'|y_1|)\times B(y_2,\epsilon'|y_1|^{1-\xi/2})\cap
B(z_1,\epsilon'|z_1|)\times B(z_2,\epsilon'|z_1|^{1-\xi/2})=\emptyset.
\end{equation}
\end{apulause}

\begin{proof}
Let
\begin{equation}
\alpha:=\frac{d(y,z)^{2/(2-\xi)}}{|y_1|},
\end{equation}

Then we have
\begin{equation}
|z_1|\le|y_1|+|y_1-z_1|\le |y_1|+d(y,z)^{2/(2-\xi)}\le (1+\alpha)|y_1|.
\end{equation}

So to prove \equref{equ-tech1}, we just have to find $\epsilon'>0$ so that
\begin{equation}
\epsilon'((1+\alpha)|y_1|)^{1-\xi/2}\le \frac 12(\alpha|y_1|)^{1-\xi/2},
\end{equation}
whenever $\alpha\ge(\epsilon'')^{2/(2-\xi)}$. By elementary calculus,
we see that this is possible.

Using similar reasoning, we see that to prove \equref{equ-tech2} we have to
find $\epsilon'>0$ so that
\begin{enumerate}
\item $\epsilon'|y_1|+\epsilon'(1+\alpha)|y_1|\le\alpha|y_1|$ and
\item $\epsilon'|y_1|^{1-\xi/2}+\epsilon'((1+\alpha)|y_1|)^{1-\xi/2}\le (\alpha|y_1|)^{1-\xi/2}$,
\end{enumerate}
when $\alpha\ge(\epsilon'')^{2/(2-\xi)}$. Again, this is possible.
\ifcmp\qed\fi
\end{proof}

\begin{propositio}\label{pro-kil}
Suppose $A\sim^\lambda \sigma(M_{n_1})\oplus...\oplus\sigma(M_{n_k})\oplus\one$
on $\R^{ld+(n-l)d}$ with $\sum_{i=1}^k (n_i-1)=l$ and
let $\epsilon''>0$ be given. Then there is $C<\infty$ such that
for $\mu\ge\epsilon''|y_1|^{1-\xi/2}$ we have
\begin{equation}
\prob^y_A(\sup_{s\le t}d(X_s,y)\ge\mu)\le C\exp\{-\frac{\mu^2}{Ct}\}.
\end{equation}
\end{propositio}

\begin{proof}
Let $\epsilon'>0$ be given by \lemref{lem-tech}.
By \corref{cor-prod}, there is $C_1<\infty$ so that if
$d(y,z)\ge\epsilon'|y_1|^{1-\xi/2}$ we have
\begin{equation}\label{equ-kil1}
K_A(t,y,z)\le C_1 t^{-\frac{ld}{2-\xi}-\frac{(n-l)d}{2}}
\exp\{-\frac{|y_1-z_1|^{2-\xi}+|y_2-z_2|^2}{C_1t}\}.
\end{equation}

Now a direct computation gives
\begin{equation}
\begin{split}
\prob^y_A(\sup_{s\le t}d&(X_s,y)\ge\mu)
\le \prob^y_A(d(X_t,y)\ge\mu/2)\\
&\quad+\prob^y_A(d(X_t,y)\le\mu/2\text{ and }
\exists s<t: d(X_s,y)=\mu)\\
&\le \prob^y_A(d(X_t,y)\ge\mu/2)\\
&\quad+\prob^y_A(\exists s<t:d(X_s,s)=\mu\text{ and }d(X_s,X_t)\ge\mu/2)\\
&\le \prob^y_A(d(X_t,y)\ge\mu/2)+\sup_{d(y,z)=\mu, s\le t}\prob^z_A(d(X_s,z)\ge\mu/2)\\
&=(*).
\end{split}
\end{equation}

By \equref{equ-tech1} of \lemref{lem-tech}, for every $z\in\R^{nd}$
with $d(y,z)=\mu$ we have
\begin{equation}
\{z':d(z,z')\le \epsilon'|z_1|^{1-\xi/2}\}\subseteq\{z':d(z,z')\le \frac\mu 2\}.
\end{equation}

A fortiori we also have
\begin{equation}
\{z':d(y,z')\le \epsilon'|y_1|^{1-\xi/2}\}\subseteq\{z':d(y,z')\le \frac\mu 2\},
\end{equation}
since there are points $z\in \R^d$ with $d(y,z)=\mu$ and $|z_1|\ge|y_1|$.

Thus by \equref{equ-kil1} we can conclude that
\begin{equation}
\begin{split}
(*)&\le C_2\int_{d(y,z)\ge\mu/2}t^{-\frac{ld}{2-\xi}-\frac{(n-l)d}2}\exp\{-\frac{|y_1-z_1|^{2-\xi}+|y_2-z_2|^2}{C_1t}\}\,dy\\
&\le C_3\int_{|y_1-z_1|^{2-\xi}\ge \mu^2}t^{-\frac{ld}{2-\xi}}
\exp\{-\frac{|y_1-z_1|^{2-\xi}}{C_1t}\}\,dy_1\\
&+ C_3\int_{|y_2-z_2|\ge \mu} t^{-\frac{(n-l)d}2}
\exp\{-\frac{|y_2-z_2|^2}{C_1t}\}\,dy_2\\
&\le C\exp\{-\frac{\mu^2}{Ct}\}.
\end{split}
\end{equation}
\ifcmp\qed\fi
\end{proof}

Now we can finish with the local estimates.

\begin{teoreema}\label{thm-loc}
Suppose that $A\sim^\lambda \sigma(M_{n_1+1})\oplus...\oplus\sigma(M_{n_k+1})\oplus\one$
on $B(0,2)\times B(0,2)$ with $l:=n_1+...+n_k<n$ and that $A$ has a heat
kernel. For any $\epsilon''\in(0,1]$ there is $C<\infty$ so that if
$y\in Q$, $0<t \le 1$ and
$\epsilon''|y_1|^{1-\xi/2}\le d(z,y)\le \frac 12$ we have
\begin{equation}
K_{A}(t,y,z)\le Ct^{-ld/(2-\xi)-(n-l)d/2}\exp\{\frac{|y_1-z_1|^{2-\xi}+|y_2-z_2|^2}{Ct}\}.
\end{equation}
Moreover, this estimate depends on $A$ only through $\lambda$, $n_1,...,n_k$
and $n$.
\end{teoreema}

\begin{remarkki}\label{rem-domains}
It is not difficult to modify the proof to take into account more general sets.
One can replace $B(0,2)\times B(0,2)$ with $U:=A\times B$ with 
$A$ and $B$ open, starlike w.r.t. origin, open and satisfying
\begin{equation}\label{equ-domains}
\bigcup_{y\in Q}\{z:d(z,y)\le\frac 12\}\subset\subset U.
\end{equation}
Similarly $Q$ can be replaced with $Q':=A\times B$ with $A$ and $B$ closed and starlike
w.r.t. origin. 

Also $d$ can be replaced with any equivalent metric. (Note in particular that \lemref{lem-tech}
is preserved under replacement by an equivalent metric with possibly a different $\epsilon'$)
\end{remarkki}

\begin{proof}
If $0< d(z,y)^2\le t$, then there is $C<\infty$ so that
\begin{equation}
1\le C\exp\{-\frac{|y_1-z_1|^{2-\xi}+|y_2-z_2|^2}{Ct}\}.
\end{equation}

Thus in view of \corref{cor-lub1} we only need to prove the claim for
$t\le d(z,y)^2\le 1$.

Let $\epsilon'>0$ be given by \lemref{lem-tech} and let $\{U_t^y\}$
be a collection of open coverings given by \proref{pro-har2}
and \remref{rem-mod} associated with this $\epsilon'$. We may assume
$\epsilon'\le\min\{\frac 12, \frac 1{2^\xi}\}$.

We want to show that  $U^t_z$ and
$B(y_1,\epsilon' t^{1/(2-\xi)})\times B(y_2,\epsilon'\sqrt{t})$
are disjoint whenever $z\not\in B(y_1,t^{1/(2-\xi)})\times
B(y_2,\sqrt{t})$. The case $|y_1|^{2-\xi}\le t \le 1$ follows easily,
since we assumed $\epsilon'\le\min\{\frac 12, \frac 1{2^\xi}\}$.
In case $0<t\le|y_1|^{2-\xi}$ we just use \lemref{lem-tech}
to conclude that
\begin{equation}
B(y_1,\epsilon' t^{1/(2-\xi)})\times B(y_2,\epsilon'\sqrt{t})
\cap B(z_1,\epsilon'|z_1|)\times B(z_2,\epsilon'|z_1|^{(2-\xi)/2})=\emptyset,
\end{equation}
whenever $d(y,z)\ge\epsilon'' |y_1|^{1-\xi/2}$.

By the proof of \corref{cor-lub1} we have
\begin{equation}
\begin{split}
t^{\frac{ld}{2-\xi}+\frac{(n-l)d}2}&\sup_{z'\in U^t_z}K_{M_{n+1}}(t,y,z')\\
&\le C_2\int_{U^t_z}dy'\,K_{M_{n+1}}(2t,x,y').
\end{split}
\end{equation}

By \proref{pro-kil} we have
\begin{equation}
\int_{U^t_z}K_{M_{n+1}}(2t,y,z)\le C_3\exp\{-\frac{|y_1-z_1|^{2-\xi}
+|y_2-z_2|^2}{C_3t}\},
\end{equation}
so we are done.
\ifcmp\qed\fi
\end{proof}

\begin{remarkki}\label{rem-estsame}
Note that the conclusion of the Theorem depends on $n_1,...,n_k$
only through $l$. In particular the estimate obtained 
above remains the same, when 
$\sigma(M_{n_1+1})\oplus...\oplus\sigma(M_{n_k+1})$ is replaced by
$\sigma(M_2)^{\oplus l}$.
\end{remarkki}

\section{Construction of the stationary state}\label{sec-stat}

In this section, we shall finally prove \thmref{thm-main2} modulo some
technicalities whose proofs are postponed until \appref{app-prc}.
To this end, we shall inductively show the following

\begin{teoreema}\label{thm-sta}
Let $\chi:\R^d\rightarrow\R$ be compactly supported and nonnegative.
Then for some $C_n<\infty$ we have
\begin{equation}
M^{-1}_{2n}(M^{-1}_{2n-2}(...(M^{-1}_2\chi\otimes \chi)...)\otimes \chi)
\le C_n \prod_{i=1}^n(1+|x_{2i-1}|)^{2-\xi-d}.
\end{equation}
\end{teoreema}

Obviously \thmref{thm-main2} follows directly from this.

The following formula is a central tool in this section.
\begin{propositio}\label{pro-bai}
Let $1\le l\in \N$. Then
\begin{equation}
\int_{\R^{ld}}d^{ld}y\,|x-y|^{2-\xi-ld}\prod_{i=1}^{l-1}(1+|y_i|)^{2-\xi-d}
\chi(y_l)\le C\prod_{i=1}^l(1+|x_i|)^{2-\xi-d}.
\end{equation}
\end{propositio}

The proof of this Proposition can be found in \appref{app-prc}.

We want to show that
\begin{equation}\label{equ-mainwant}
\int_{\R^{(2n-1)d}}G_{M_{2n}}(x,y)\prod_{i=1}^{n-1}(1+|y_{2i-1}|)^{2-\xi-d}
\chi(y_{2n-1})\,dy\le C\prod_{i=1}^n(1+|x_i|)^{2-\xi-d}.
\end{equation}

We find finitely many sets $\{A_i\}_{i=1}^k$ so that together
with $\{(x,y)\in \R^{(2n-1)d}\times\R^{(2n-1)d}:|y|\ge\rho|x|\}$ they
cover $\R^{(2n-1)d}\times\R^{(2n-1)d}$. Let $A^x_i:=\{x+y:(x,y)\in\R^{nd}\}$. 

We shall write the above integral as
\begin{equation}\label{equ-splitting}
\int_{\R^{(2n-1)d}}=\int_{|x-y|\ge\rho|x|}+\sum_{j=0}^{k(x)}\int_{A^x_j\setminus A^x_{j-1}}
\end{equation}
and then prove the desired estimate of \equref{equ-mainwant} separately
for each term of the right-hand side.

We apologise the reader for bouncing around with using $2n$ and $n+1$, but for the
moment $n+1$ is more convenient.

We will first reduce everything to investigation of operators $\sigma(M_2)^{\oplus l}$ using \remref{rem-estsame}.
What we mean by this is the following: Let
\begin{equation}
E_C(t,x,y):=
\begin{cases}
C|x|^{-\frac{\xi d}{2}}t^{-\frac d2}\exp\left\{-\frac{|x|^{-\xi}|x-y|^2}{Ct}\right\}&\text{if $|y|<\frac{|x|}2$}\\
Ct^{-\frac{d}{2-\xi}}\exp\left\{-\frac{|x-y|^{2-\xi}}{Ct}\right\}&\text{if $|y|\ge\frac{|x|}2$}\\
\end{cases}
\end{equation}
and let
\begin{equation}
E^n_C(x,y):=\int_0^\infty dt\,\prod_{i=1}^n E_C(t,x_i,y_i).
\end{equation}

We want to find $C<\infty$ and a finite covering
$\{A_i\}_{i=1}^m$ for $\R^{nd}\times\R^{nd}$ so that for every $i\in\{1,...,m\}$
there is $L_i\in\mathcal L'_{n+1}$ so that
\begin{equation}
G_{M_n^{L_i}}(t,x,x+y)\le E^n_C(t,x,x+y)
\end{equation}
whenever $(x,y)\in LA_i$.

Then the proof of \equref{equ-mainwant} is reduced to the investigation of $E_C(t,x,y)$ (which is just
the natural estimate for $\sigma(M_2)^{\oplus n}$).

We'll first define $A_i$'s for
symbols $A\sim\bigoplus_{i=1}^k\sigma(M_{n_i+1})\oplus\one$ on $B(0,2)\times B(0,2)$
by induction on $l:=\sum_{i=1}^k$ and then use these to define $A_i$'s for $\sigma(M_n)$.
In this local case we just cover $\overline{B}(0,2)\times \overline{B}(0,2)\times \overline{B}(0,\epsilon)$.

So suppose we have just a uniformly elliptic operator $A$ on $B(0,2)$. Then we just take one
set $A_1:=\{(x,y):x\in B(0,1)\text{ and }|x-y|<\frac 12\}$. Next suppose all the cases $l'<l$
have been handled. Then by induction hypothesis and compactness of $\Sph^{ld-1}\times \overline{B}(0,1)$
there exists a finite set $\{x_1,...,x_m\}$ of $\Sph^{ld-1}\times \overline{B}(0,1)$ so that 
there are affine transformations $K_1,...,K_m$ so that each $K_j$ sends $x_j$ to $0$ and 
$A^{K_j}\sim \bigoplus_{i=1}^{k_i}\sigma(M_{n^j_i+1})\oplus\one$ with $l_j=\sum_{i=1}^{k_i} n^j_i<l$
on $B(0,2)\times B(0,2)$. 

Since $l_j<l$, there are $\{A_i\}_{i=1}^k$ so that
$\Sph^{ld-1}\times\overline{B}(0,1)\times\overline{B}(0,\epsilon)$ gets covered by them and each $A_i$
is just $(L^{-1}_j)^{\oplus 2} A$ for some associated $A$ given for $A^{K_j}$.

Moreover the linear part $L_j$ of $K_j$ is of the form
\begin{equation}
L_j:=
\begin{pmatrix}
M_j &0\\
0 &\one
\end{pmatrix},
\end{equation}
where $M_j$ is a $ld\times ld$-matrix. So there is a neighbourhood 
\begin{equation}
B_\epsilon:=\{(x,y):x\in\Sph^{ld-1}\times \overline{B}(0,1)\text{ and }|x-y|<\epsilon\}
\end{equation}
so that on $B_\epsilon\subseteq \bigcup_{i=1}^m U_i$ everything is under control.

Let's define the set $\tilde{A}_i$ as follows:
\begin{multline}
\tilde{A}_i:=\{((rx_1,x_2),(ry_2,r^{1-\xi/2}y_2)):\\
(x,y)\in A_i, x\in\Sph^{ld-1}\times\overline{B}(0,1)\text{ and }r\in(0,1]\}. 
\end{multline}

Clearly
there is $\epsilon''>0$ so that $\{\tilde{A}_i\}_{i=1}^m$ together with (see again \thmref{thm-loc})
\begin{equation}
\{\epsilon''|y_1|^{1-\xi/2}\le d(z,y)\le \frac 12\}
\end{equation}
cover 
\begin{equation}
\{(x,y): x\in\overline{B}(0,1)\times\overline{B}(0,1)\text{ and }|x-y|<\epsilon\}
\end{equation}
for some $\epsilon>0$.

On this last set $A$ clearly ``behaves as'' the heat kernel of $\sigma(M_2)^{\oplus l}\oplus\one$,
so we have to prove the same for $A^{L_i}$ on $L_i \tilde{A}_i$. This is a rather easy scaling argument:
Pick $\lambda>0$ so that $A\sim^\lambda\bigoplus_{i=1}^k\sigma(M_{n_i+1})\oplus\one$.
Let $y\in \overline{B}(0,1)\times \overline{B}(0,1)$ and denote
$x^y:=(|y_1|^{-1}x_1,|y_1|^{\xi/2-1}(y_2-x_2)+x_2)$. Define
\begin{equation}
B^y_{ij}(z^y):=
\begin{cases}
|y_1|^{\xi}A_{ij}(z)&\text{if $1\le i, j\le ld$}\\
|y_1|^{\xi/2}A_{ij}(z)&\text{if $1\le i\le ld < j\le nd$
or }\\&\text{$1\le j\le ld< i\le nd$}\\
A_{ij}(z)&\text{if $ld< i,j\le nd$}.
\end{cases}
\end{equation}

A straightforward computation shows that $B^y\sim^\lambda\bigoplus_{i=1}^k\sigma(M_{n_i+1})\oplus\one$.
By dimensional analysis
\begin{equation}
G_{A}(y,z)=|y_1|^{2-\xi-ld-(1-\xi/2)(n-l)d}G_{B^y}(y^y,z^y).
\end{equation}

Therefore, since the same scaling property holds for $E^n_C$, we can conclude that
\begin{equation}
G_{A^{L_i}}(y,z)\le E_C^n(y,z)
\end{equation}
on whole of $L_i \tilde{A_i}$.

Finally for $\sigma(M_n)$ we just cover $\Sph^{nd-1}\times \overline{B}(0,\rho)$ by the
sets described above and conify these. Now if $\sigma(M_{n+1})^L$
``behaves as'' $\sigma(M_2)^{\oplus l}\oplus\one$ on $LA$, then by scaling it ``behaves as''
$\sigma(M_2)^{\oplus l}\oplus\sigma(M_2)^{\oplus (n-l)}=\sigma(M_2)^{\oplus n}$ on $\mathcal C LA$,
where $\mathcal C LA$ denotes the conification of $LA$.

So we have reduced \equref{equ-mainwant} to proving
\begin{multline}\label{equ-mainwant2}
\int_{\R^{(2n-1)d}}E^{2n-1}_C(x,y)\prod_{i=1}^{n-1}(1+|Ly_{2i-1}|)^{2-\xi-d}\chi(Ly_{2n-1})\,dy\\
\le C'\prod_{i=1}^n(1+|Lx_i|)^{2-\xi-d}
\end{multline}
for arbitrary $L\in \mathcal L'_{2n}$ and arbitrary $C>0$.

To this end, we split the domain of integration in \equref{equ-mainwant2} into parts and
prove it separately for these parts.

Define the sets $B^x_j$ as follows: Assume first that $\max\{|x_i|:1\le i\le n\}=1$.
(We then just simply let $B^x_j:=r B^{x/r}_j$ if $\max\{|x_i|:1\le i\le n\}=r$.

By symmetry we may assume $|x_1|\le...\le|x_n|=1$.
Let \ifcmp{\linebreak[4]}\fi $k(x):=\#(\{|x_1|,...,|x_n|\}\setminus\{0\})$ (i.e. the number of
distinct strictly positive numbers) and let $\ell$ be defined by
\begin{equation}
\begin{split}
0&<|x_{\ell^x_1}|=...=|x_{\ell^x_2-1}|<|x_{\ell^x_2}|=...=|x_{\ell^x_3-1}|<...\\
&<|x_{\ell^x_{k(x)}}|=...=|x_n|=1
\end{split}
\end{equation}
with $\ell^x_1$ being the smallest integer so that $|x_{\ell^x_1}|>0$. Let
$r^x_j:=|x_{\ell^x_j}|$.

For every $x$ with
$k(x)>1$, we define $\tilde{x}$ as follows:
\begin{enumerate}
\item $\tilde{x}_i=x_i/r^x_{k(x)-1}$ for $1\le i< \ell^x_{k(x)}$ and
\item $\tilde{x}_i=x_i=1$ for $\ell^x_{k(x)}\le i\le n$.
\end{enumerate}

Clearly for such $x$, $k(\tilde{x})=k(x)-1$. We first give the sets
$B^x_j$ inductively in terms of $k(x)$ and then explicitly.
First of all, for all our $x$ let
\begin{equation}
B^x_{k(x)}:=\{y\in\R^{nd}:|y_i-x_i|\le\frac 12\text{ for every }i\in\{1,\cdots,n\}.
\end{equation}

In particular, for $k(x)=1$ everything is done. If $k(x)>1$ and $B^y_j$ has been
defined for $y$ with $k(y)<k(x)$, we just
translate $B^{\tilde{x}}_j$ on top of $x$ and scale it by $r^x_{k(x)-1}$
in the first $\ell^x_{k(x)}-1$ coordinates and by a factor of $(r^x_{k(x)-1})^{1-\xi/2}$
in the rest of the coordinates. In plain formulese, this is
\begin{equation}\label{equ-scale}
\begin{split}
B^{x}_j:=\{(r y_1,&...,r y_{\ell^x_{k(x)}-1},
x_{\ell^x_{k(x)}}+r^{1-\xi/2}(y_{\ell^x_{k(x)}}
-x_{\ell^x_{k(x)}}),\\&...,x_n+r^{1-\xi/2}(y_n-x_n)):y\in B^{\tilde{x}}_j\}.
\end{split}
\end{equation}

Thus, explicitly, we have (denoting $\ell^x_{k(x)+1}=n+1$)
\begin{equation}\label{equ-explform}
\begin{split}
B^{x}_j:=\{y\in\R^{nd}:&\forall i<\ell^x_{j+1}:|x_i-y_i|\le
\frac 12|x_{\ell^x_j}|\text{ and}\\
&\forall l\in\{j+1,...,k(x)\}\forall i\in\{\ell^x_l,...,\ell^x_{l+1}-1\}:\\
&|x_i-y_i|\le\frac 12|x_{\ell^x_j}|^{1-\xi/2}|x_{\ell^x_{l}}|^{\xi/2}\}.
\end{split}
\end{equation}

For technical reasons related to the fact that the symmetry group of
$\sigma(M_{n+1})$ (i.e. $\mathcal L'_{n+1}$) is rather different
from the one of $\sigma(M_2)^{\oplus n}$ we have to modify our
covering $\{B^x_j\}$ a bit, since with our current covering
\itemref{item:2} of \lemref{lem-axjs3} would not be true. (It is
true however for such $L$ for which $\{Ly_l=0\}=\{y_{l'}=0\}$
for some other $l'$).

First of all, we can concentrate our investigation
to a conical neighbourhood $\mathcal C$ of the degeneration set,
since outside such neighbourhood for $|x-y|\le\rho|x|$ we have
\begin{equation}\label{equ-reduce}
E^n_C(x,y)\le C |x|^{-\xi}|x-y|^{2-nd} \le C'|x-y|^{2-\xi-nd},
\end{equation}
and this is sufficient by the computation in Phase 1 of the proof
of \thmref{thm-sta} below.

Therefore, we pick our conical neighbourhood $\mathcal C$
and $\rho>0$ so that the set
\begin{equation}
\bigcup_{x\in \mathcal C} B(x,3\rho|x|)
\end{equation}
does not contain any of the degeneration points of $\sigma(M_{n+1})$
that are not degeneration points of $\sigma(M_{n+1})$. Then we just
intersect each $B^x_j$ for $x\in\mathcal C$ with
\begin{equation}
\bigcup_{x\in \mathcal C} B(x,2\rho|x|)
\end{equation}
thus forcing \itemref{item:2} of \lemref{lem-axjs3} to be true. 

After this small diversion, we shall now list some basic properties of $E^n_C$ on these sets
needed to establish \equref{equ-mainwant}. The statements of Lemmata \ref{lem-axjs1},
\ref{lem-axjs2} and \ref{lem-axjs3} clearly scale if for general $x$ we define
$B^x_j$ to be $rB^{x/r}_j$, where $r:=\max\{|x_1|,...,|x_n|\}$. So we can assume
that $r=1$ in the following proofs.

\begin{apulause}\label{lem-axjs1}
For every
$x\in\mathcal C$ with $|x_1|\le\cdots\le|x_n|$, $j\in\{1,...,k(x)\}$ and
$y\in\mathcal B^x_j$ there are some positive numbers $a_{\ell^x_j},...,a_n$ so that
\begin{equation}
  \label{eq:1}
\begin{split}
E^n_C(x,y)\le &C'\bigl(\prod_{i=\ell^x_j}^{n} a_i^{-\frac{\xi d}2}\bigr)
\bigl(\sum_{i=1}^{\ell^x_j-1}|x_i-y_i|^{2-\xi}+\\
&\qquad+\sum_{i=\ell^x_j}^{n}a_i^{-\xi}|(x_i-y_i|^2\bigr)^{1-\frac{ld}{2-\xi}-
\frac{(n-l)d}2}
\end{split}
\end{equation}
on $B^x_j\setminus B^x_{j-1}$.
\end{apulause}

\begin{proof}
This is just a straightforward computation by induction on $k(x)$.
\end{proof}

\begin{apulause}\label{lem-axjs2}
There is $C'<\infty$ such that
\begin{equation}
  \label{eq:2}
  \int_{B^x_j\setminus B^x_{j-1}}E^n_C(x,y)\,dy\le C'(r^x_j)^{2-\xi}.
\end{equation}
\end{apulause}

\begin{proof}
Again by induction on $k(x)$.
\end{proof}

\begin{apulause}\label{lem-axjs3}
  Let $L\in \mathcal L'_{n+1}$ ($\mathcal L'_{n+1}$ was defined in \equref{equ-mcl}).
  Then there is $C'<\infty$ such
  that for every $x\in\mathcal C$, $j\in\{1,\cdots,k(x)\}$ and $l\in\{1,...,n\}$ the following hold:
  \begin{enumerate}
  \item \label{item:1}If $\{y_i=0:1\le i \le \ell^x_j-1\}\subseteq\{Ly_l=0\}$, then $|Lx_l|\le C' r^x_j$.
  \item \label{item:2}If $\{y_i=0:1\le i \le \ell^x_j-1\}\not\subseteq\{Ly_l=0\}$, then $B^x_j\subseteq \{|Ly_l|\ge C^{\prime -1}|Lx_l|\}$.
  \end{enumerate}
\end{apulause}

\begin{proof}
As before, without loss of generality we may assume that \linebreak[4] $\max\{|x_1|,...,|x_n|\}=1$ and
that $|x_1|\le\cdots\le|x_n|$.

First we handle \itemref{item:1}. Basically it says that if $Ly_l$ can be expressed as a linear
combination of $y_i$'s with $1\le i \le \ell^x_j-1$, then $|Lx_l|$ is small. Since
$\max_{L\in\mathcal L'_n+1}||L||\le C$ for some $C<\infty$, it
suffices to show that in case of $L=\one$ we have $|x_l|\le r^x_j$.
But this is immediate from the fact that $|x_l|\le |x_{\ell^x_{j-1}}|\le |x_{\ell^x_j}|=r^x_j$.

Also in the case of \itemref{item:2}, we can immediately restrict our attention to the case
$L=\one$. This is then immediate using \equref{equ-explform}.
\end{proof}

\begin{proof}(of \thmref{thm-sta})
As was said before, it suffices to prove \equref{equ-mainwant2} for
$x\in \mathcal C$, since for $x\not\in\mathcal C$ the whole thing
reduces to Phase 1 below using \equref{equ-reduce}.

Without loss of generality, we may assume that the support of $\chi$
is so small that if \itemref{item:2} applies to $y_{2n-1}$, then whenever
$r^x_j\ge 1$, we have
\begin{equation}\label{equ-suppsmall}
L[\R^{(2n-2)d}\times \supp\chi]\subseteq\{|Ly_{2n-1}|<C^{\prime -1}|Lx_{2n-1}|\}.
\end{equation}

Our proof goes as follows. First we split the domain of integration into
parts
\commentout{
\begin{equation}
\int_{\R^{(2n-1)d}}=\int_{|x-y|\ge\rho|x|}+\sum_{j=0}^{k(x)}\int_{A^x_j\setminus A^x_{j-1}}
\end{equation}
}
and then we proceed in three phases:
\begin{enumerate}
\item Phase 1: Handle the integral $\int_{|x-y|\ge\frac \rho|x|}$.
\item Phase 2: Handle the integrals $\int_{B^x_j\setminus B^x_{j-1}}$ with $r^x_j\le 1$ and
\item Phase 3: Handle the integrals $\int_{B^x_j\setminus B^x_{j-1}}$ with $r^x_j\ge 1$.
\end{enumerate}

First, phase 1: 
We know that for $|x-y|\ge\frac 12|x|$ we have 
\begin{equation}
E^{2n-1}_C(x,y)\le C_1(\sum_{i=1}^{2n-1}|Lx_i-Ly_i|)^{2-\xi-2(n-1)d},
\end{equation}
so we can conclude that
\begin{equation}
\begin{split}
\int_{|x-y|\ge \rho|x|}&d^{(2n-1)d}y\,E^{2n-1}_C(x,y)
\prod_{i=1}^{n-1}(1+|Ly_{2i-1}|)^{2-\xi-d}\chi(|Ly_{2n-1}|)\\
&\le C_1\int_{\R^{(2n-1)d}}d^{(2n-1)d}y\,(\sum_{i=1}^{2n-1}|Lx_i-Ly_i|)^{2-\xi-(2n-1)d}\cdot\\
&\qquad\cdot\prod_{i=1}^{n-1}(1+|Ly_{2i-1}|)^{2-\xi-d}\chi(|Ly_{2n-1}|)=:(*)\\
\end{split}
\end{equation}

By a change of variables we see that
\begin{equation}\label{equ-chav}
\begin{split}
\int_{\R^{(n-1)d}}&\prod_{i=1}^{n-1}d^d y_{2i}\,(\sum_{i=1}^{2n-1}|x_i-y_i|)^{2-\xi-(2n-1)d}\\
&=(\sum_{i=1}^n|x_{2i-1}-y_{2i-1}|)^{2-\xi-nd}\int_{\R^{(n-1)d}}\prod_{i=1}^{n-1}d^d z_{2i}\,\cdot\\
&\qquad\cdot(1+\sum_{i=1}^{n-1}|z_{2_i}|)^{2-\xi-(2n-1)d},\\
\end{split}
\end{equation}
so we can conclude that
\begin{equation}
\begin{split}
(*)&\le C_1\int_{\R^{nd}}\prod_{i=1}^nd^dy_{2i-1}\,(\sum_{i=1}^n|Lx_{2i-1}-Ly_{2i-1}|)^{2-\xi-nd}\cdot\\&\qquad\cdot\prod_{i=1}^{n-1}(1+|Ly_{2i-1}|)^{2-\xi-d}\chi(|Ly_{2n-1}|)=:(*^2).
\end{split}
\end{equation}

By \proref{pro-bai}, we have
\begin{equation}
(*^2)\le C_2\prod_{i=1}^n(1+|Lx_{2i-1}|)^{2-\xi-d}.
\end{equation}

Next, some initial preparation for phases 2 and 3: Let $x$ and $j\le k(x)$ 
be given and let $U:=\{1,3,...,2n-1\}$,
let $U_1$ be the set of those $l\in U$ for which \itemref{item:2} of \lemref{lem-axjs3} applies
and let $U_2:=U\setminus U_1$.

Then, phase 2: so suppose $r^x_j\le 1$. Then by
\lemref{lem-axjs2} we have
\begin{equation}
\begin{split}
\int_{y\in B^x_j\setminus B^x_{j-1}}&d^{(n-1)d}y\,E^{2n-1}_C(x,y)
\prod_{i=1}^{n-1}(1+|Ly_{2i-1}|)^{2-\xi-d}\chi(|Ly_{2n-1}|)\\
&\le C_3\sup_{y\in B^x_j\setminus B^x_{j-1}}\prod_{i=1}^n(1+|Ly_{2n-i}|)^{2-\xi-d}=:(*^3)
\end{split}
\end{equation}

Since $(1+|y_i|)^{2-\xi-d}\le 1$ for any $i\in U$, we can conclude that
\begin{equation}
(*^3)\le C_3\sup_{y\in B^x_j\setminus B^x_{j-1}}\prod_{i\in U_2}(1+|Ly_i|)^{2-\xi-d}:=(*^4)
\end{equation}

By \itemref{item:2} of \lemref{lem-axjs3} we have
\begin{equation}
(*^4)\le C_4\prod_{i\in U_2}(1+|Lx_i|)^{2-\xi-d}=:(*^5)
\end{equation}

Since by \itemref{item:1} of \lemref{lem-axjs3} we have $|Lx_i|\le C'$ for
$i\in U_1$ we can finally conclude that
\begin{equation}
(*^5)\le C_5\prod_{i=1}^{n}(1+|Lx_{2i-1}|)^{2-\xi-d}.
\end{equation}

Finally, phase 3: If $r^x_j\ge 1$ and $2n-1\in U_2$, then 
by \equref{equ-suppsmall} we have
\begin{equation}\label{equ-support}
L[\R^{(2n-2)d}\times \supp\ \chi]\cap B^x_j=\emptyset.
\end{equation}
by \itemref{item:2} of \lemref{lem-axjs3}
and thus in this case we have
\begin{equation}
\int_{y\in B^x_j}d^{(n-1)d}y\,E^{2n-1}_C(x,y)
\prod_{i=1}^{n-1}(1+|Ly_{2i-1}|)^{2-\xi-d}\chi(|Ly_{2n-1}|)=0.
\end{equation}

So we may assume $2n-1\in U_1$. By \itemref{item:2} of \lemref{lem-axjs3} 
we have
\begin{equation}
(1+|Ly_i|)^{2-\xi-d}\le C_6(1+|Lx_i|)^{2-\xi-d}
\end{equation}
for $i\in U_2$. Therefore
\begin{equation}
\begin{split}
\int_{y\in B^x_j\setminus B^x_{j-1}}&d^{(2n-1)d}y\,E^{2n-1}_C(x,y)
\prod_{i=1}^n(1+|Ly_{2i-1}|)^{2-\xi-d}\\
&\le C_7\prod_{i\in U_2}(1+|Lx_i|)^{2-\xi-d}\int_{y\in B^x_j\setminus B^x_{j-1}}
d^{(2n-1)d}y\,
E^n_C(x-y)\cdot\\
&\qquad\cdot\prod_{i\in U_1\setminus\{2n-1\}}(1+|Ly_i|)^{2-\xi-d}\chi(Ly_{2n-1})
=(*^6).
\end{split}
\end{equation}

Writing $l':=2n-1-l$ we get
\begin{equation}
\begin{split}
\int_{y\in B^x_j\setminus B^x_{j-1}}&d^{(2n-1)d}y\,E^{2n-1}_C(x-y)\prod_{i\in U_1\setminus\{2n-1\}}(1+|Ly_i|)^{2-\xi-d}\chi(Ly_{2n-1})\\
&\le C_8\int_{\R^{(2n-1)d}}d^{(2n-1)d}y\,\bigl(\prod_{i=l+1}^{n} a_i^{-\frac{\xi d}2}\bigr)
\bigl(\sum_{i=1}^{l}|x_i-y_i|^{2-\xi}+\\
&\qquad+\sum_{i=l+1}^{n}a_i^{-\xi}|x_i-y_i|^2\bigr)^{1-\frac{ld}{2-\xi}-
\frac{l'd}2}\cdot\\
&\qquad\qquad\cdot\prod_{i\in U_1\setminus\{2n-1\}}(1+|Ly_i|)^{2-\xi-d}\chi(Ly_{2n-1})=(*^7),
\end{split}
\end{equation}

Note that by the definition of $U_1$ 
we have that in the expression
\begin{equation}
\prod_{i\in U_1\setminus\{2n-1\}}(1+|Ly_i|)^{2-\xi-d}\chi(Ly_{2n-1})
\end{equation}
depends only on the variables $y_1,...,y_l$.

Using a similar change of variables as in \equref{equ-chav}, we get
\begin{equation}
\begin{split}
(*^7)&\le C_8\int_{\R^{ld}}\prod_{i=1}^ld^d y_i\,(\sum_{i=1}^l
|x_i-y_i|^{2-\xi})^{1-\frac{ld}{2-\xi}}\cdot\\
&\qquad\cdot\prod_{i\in U_1\setminus\{2n-1\}}(1+|Ly_i|)^{2-\xi-d}\chi(Ly_{2n-1})
\int_{\R^{l'd}}
\prod_{i=l+1}^{2n-1} d^d y_i \cdot\\
&\qquad\cdot\bigl(\prod_{i=l+1}^{2n-1} a_i^{-\frac{\xi d}2}\bigr)(1+\sum_{i=l+1}^{2n-1}a_i^{-\xi}|x_i-y_i|^2)^{1-\frac{ld}{2-\xi}-\frac{l'd}2}=(*^8).\\
\end{split}
\end{equation}

By substituting $y'_i=a_i^{-\xi/2}y_i$ we see that the last integral
is $\le C_9$.

Therefore
\begin{equation}
\begin{split}
(*^8)&\le C_{10}\int_{\R^{ld}}\prod_{i=1}^ld^d y_i\,(\sum_{i=1}^l
|x_i-y_i|^{2-\xi})^{1-\frac{ld}{2-\xi}}\cdot\\&\qquad\cdot
\prod_{i\in U_1\setminus\{2n-1\}}(1+|Ly_i|)^{2-\xi-d}\chi(Ly_{2n-1})=:(*^9).
\end{split}
\end{equation}

Noticing that $\sum_{i=1}^l|x_i-y_i|^{2-\xi}$ is essentially just
$|(x_1-y_1,...,x_l-y_l)|^{2-\xi}$ for estimation purposes, using
a similar change-of-variables argument as before, we can conclude
that
\begin{equation}
(*^9)\le C_{11}\prod_{i\in U_1}(1+|Lx_i|)^{2-\xi-d},
\end{equation}
so
\begin{equation}
(*^6)\le C_{12}\prod_{i=1}^n(1+|Lx_{2i-1}|)^{2-\xi-d}.
\end{equation}
\ifcmp\qed\fi
\end{proof}

\ifcmp
\else
\section{Acknowledgements}
\acknowledgementstext
\fi

\appendix

\section{Poincar\'e and Sobolev inequalities}\label{app-poin}
The following Theorem was proved in \cite{CW}.

\begin{teoreema}\label{thm-poin}
Let $q>2$ and let $w_1$ and $w_2$ be two weights on $\R^n$ and suppose that $w_1$ is $A_2$ and that $w_2$ is doubling. Suppose also
that for all balls $B'$ and $B$ with $B'\subseteq 2B$
\begin{equation}
\biggl(\frac{|B'|}{|B|}\biggr)^{1/n}\biggl(\frac{w_2(B')}{w_2(B)}\biggr)^{1/q}
\le c\biggl(\frac{w_1(B')}{w_1(B)}\biggr)^{1/2}
\end{equation}
with $c$ independent of the balls.

Then the Poincar\'e and Sobolev inequalities hold for $w_1$, $w_2$ with $q$.
\end{teoreema}

So in order to conclude that the Harnack inequality holds for
$M_n$, it suffices to check the assumptions of above Theorem
with $w_1=d(x,F)^\xi$ with $F$ be a finite union of vector subspaces of $\R^n$ or just $\{0\}$ and $w_2$ either $|x|^\xi$ or $1$.

\begin{apulause}\label{lem-beh}
Let $F$ be a finite union of vector subspaces of $\R^n$ or just $\{0\}$.
Suppose $\xi>-n$. Then $w_\xi(x):=d(x,F)^\xi$ satisfies the following:
There is a constant $C<\infty$ such that for every $x\in\R^n$ we have
\begin{enumerate}
\item If $0<r<\frac{d(x,F)}{2}$, then $\frac{1}{C}d(x,F)^\xi r^n\le
      w_\xi(B(x,r))\le C d(x,F)^\xi r^n$ and
\item If $r\ge\frac{d(x,F)}{2}$, then $\frac{1}{C}r^{n+\xi}\le
      w_\xi(B(x,r))\le C r^{n+\xi}$.
\end{enumerate}
\end{apulause}

\begin{proof}
Since for $y\in B(x,r)\subseteq B(x,\frac{d(x,F)}{2})$ we have
$\bigl(\frac{d(x,F)}{2}\bigr)^\xi\le w_\xi(y)\le\bigl(\frac{3d(x,F)}{2}\bigr)^\xi$, the first estimate follows.

For the second estimate, since $w_\xi(x,r)=|x|^{n+\xi}w_\xi(\hat x,\frac r{|x|})$ we see that by scaling it suffices to prove the
inequality for $x\in \Sph^{n-1}$. To conclude the proof, it suffices
to prove that
\begin{equation}
0<\lim_{r\rightarrow\infty}\sup_{x\in\Sph^{n-1}}
\frac{1}{r^{n+\xi}}w_\xi(B(x,r))=\lim_{r\rightarrow\infty}\inf_{x\in\Sph^{n-1}}
\frac{1}{r^{n+\xi}}w_\xi(B(x,r))<\infty.
\end{equation} The computation is omitted.
\ifcmp\qed\fi
\end{proof}

Naturally, the choice of the borderline at $\frac{d(x,F)}{2}$ was arbitrary.
We can and will put the borderline at $\epsilon d(x,F)$ with $\epsilon\in(0,1)$
depending on the situation.

\begin{propositio}\label{pro-poin}
Let $0<\xi<2$ and let $F$ be a finite union of vector subspaces of $\R^n$
of codimension $\ge 2$ or just $\{0\}$.
Let $w_1=C_1 d(x,F)^\xi$ and $w_2=C_2 |x|^\xi$ with $0<C_1,C_2<\infty$.
Then the Poincar\'e and Sobolev inequalities hold for $w_1$, $w_2$ with
$q:=\frac{2n}{n+\xi-2}$ and for $w_1$, $1$ with $q$.
\end{propositio}
\begin{proof}
By \lemref{lem-beh} both $w_1$ and $w_2$ are $A_2$, so it suffices
to prove the scaling assumption in \thmref{thm-poin} with $q$.
Now \lemref{lem-beh} implies that there is a constant $C<\infty$ such that
for every $x\in \R^d$ and $r>0$ and every ball $B':=B(x',r')\subseteq B(x,2r)$
we have
\begin{enumerate}
\item If $0<r<\frac{d(x,F)}{4}$, then
$C^{-1}|x|^\xi r^{\prime n}\le w(B')\le C|x|^\xi r^{\prime n}$.
\item If $\frac{d(x,F)}{4}<r$, then
$C^{-1}r^{\prime n+\xi}\le w(B')\le Cr^\xi r^{\prime n}$.
\end{enumerate}
Here $w$ stood for either $w_1$ or $w_2$. Thus we have for some $C<\infty$
the following:
\begin{enumerate}
\item If $0<r<\frac{d(x,F)}{4}$, then
\begin{equation}
C^{-1}\biggl(\frac{r^{\prime n}}{r^n}\biggr)\le\biggl(\frac{w(B')}{w(B)}\biggr)\le C\biggl(\frac{r^{\prime n}}{r^n}\biggr).
\end{equation}
\item If $\frac{d(x,F)}{4}<r$, then
\begin{equation}
C^{-1}\biggl(\frac{r^{\prime n+\xi}}{r^{n+\xi}}\biggr)\le\biggl(\frac{w(B')}{w(B)}\biggr)\le C\biggl(\frac{r^{\prime n}}{r^n}\biggr)
\end{equation}
\end{enumerate}
Therefore, the claim reduces to finding $C<\infty$ such that for every $r$ and
$r'$ with $r'\le 2r$ we have:
\begin{enumerate}
\item If $0<r<\frac{d(x,F)}{4}$, then
\begin{equation}
\biggl(\frac{r'}{r}\biggr)\biggl(\frac{r^{\prime n}}{r^n}\biggr)^{\frac{n-2+\xi}{2n}}\le C\biggl(\frac{r^{\prime n}}{r^n}\biggr)^{1/2}
\end{equation}
and
\begin{equation}
\biggl(\frac{r'}{r}\biggr)^{\frac{n+\xi}{2}}\le C\biggl(\frac{r'}{r}\biggr)^{\frac{n}{2}}
\end{equation}
\item If $\frac{d(x,F)}{4}<r$, then
\begin{equation}
\biggl(\frac{r'}{r}\biggr)\biggl(\frac{r^{\prime n}}{r^n}\biggr)^{\frac{n-2+\xi}{2n}}\le C\biggl(\frac{r^{\prime n+\xi}}{r^{n+\xi}}\biggr)^{1/2}
\end{equation}
and
\begin{equation}
\biggl(\frac{r'}{r}\biggr)^{\frac{n+\xi}{2}}\le C\biggl(\frac{r'}{r}\biggr)^{\frac{n+\xi}{2}}
\end{equation}
\end{enumerate}
Obviously, such a $C$ exists, so our claim has been proved.
\ifcmp\qed\fi
\end{proof}

\section{Proofs for \secref{subsec-lem}}\label{app-lemm}

\begin{proof} (of \proref{pro-don})
Let $C_1:=\inf\{\langle \mathbf v,\sigma(M_{n+1})(\mathbf x)\mathbf v\rangle:
|\mathbf x|=|\mathbf v|=1\text{ and }\mathbf x\in A\}$ and
$C_2:=\sup\{\langle \mathbf v,\sigma(M_{n+1})(\mathbf x)\mathbf v\rangle:
|\mathbf x|=|\mathbf v|=1\text{ and }\mathbf x\in A\}$. Since $A$ is conical
with $A\cap S^{nd-1}$ compact and disjoint from the degeneration
set, we have $C_1>0$.

For $\mathbf x\in A$ we have
\begin{equation}
\begin{split}
C_1\sum_{i=1}^{n}|x_i|^\xi|v_i|^2
&\le C_1\sum_{i=1}^n|\mathbf x|^\xi|v_i|^2\\
&= C_1|\mathbf x|^\xi|\mathbf v|^2\\
&\le \sigma(M_n)\\
&\le C_2|\mathbf x|^\xi|\mathbf v|^2\\
&= C_2\sum_{i=1}^n|\mathbf x|^\xi|v_i|^2\\
&\le C_2\bigl(\frac{\sqrt n}\epsilon\bigr)^\xi\sum_{i=1}^n|x_i|^\xi|v_i|^2,
\end{split}
\end{equation}
where the last inequality follows from the fact that
\begin{equation}
\begin{split}
|\mathbf x|^\xi&=(\sum_{i=1}^n|x_i|^2)^{\xi/2}\\
&\le n^{\xi/2}\max\{|x_i|^\xi:1\le i\le n\})\\
&\le \bigl(\frac{\sqrt n}{\epsilon}\bigr)^\xi
\min\{|x_i|^\xi:1\le i\le n\}\\
&\le\bigl(\frac{\sqrt n}{\epsilon}\bigr)^\xi|x_i|^\xi.
\end{split}
\end{equation}
\ifcmp\qed\fi
\end{proof}

\begin{proof}(of \lemref{lem-cro1})
We write $|\langle v_i,(d(x_i+x_{i+1})-d(x_i)-d(x_{i+1}))v_{i+1}\rangle|
\le|\langle v_i,(d(x_i+x_{i+1})-d(x_{i+1}))v_{i+1}\rangle|+
|\langle v_i,d(x_i)v_{i+1}\rangle|$ and estimate the two terms separately.

Since $d$ is differentiable in the ball
$\overline B(x_{i+1},\frac{1}{2}|x_{i+1}|)$ a simple application of the
mean value theorem of elementary calculus gives
\begin{equation}
\begin{split}
|\langle v_i,(d(x_i+x_{i+1})-&d(x_{i+1}))v_{i+1}\rangle|\\
&\le\sup_{0\le r\le 1}\langle v_i,(x_i\cdot\nabla)d(x_{i+1}+r x_i)v_{i+1}\rangle\\
&\le\sup_{\frac{1}{2}\le |y|\le\frac{3}{2}}|\langle \hat{v_i},(\hat{x_i}\cdot\nabla)d(y)
\hat{v}_{i+1}\rangle||x_i||x_{i+1}|^{\xi-1}|v_i||v_{i+1}|\\
&:=C|x_i||x_{i+1}|^{\xi-1}|v_i||v_{i+1}|\\
&=C\bigl(\frac{|x_i|}{|x_{i+1}|}\bigr)^{1-\xi/2}|x_i|^{\xi/2}|x_{i+1}|^{\xi/2}
|v_i||v_{i+1}|\\
&\le\frac C2\bigl(\frac{|x_i|}{|x_{i+1}|}\bigr)^{1-\xi/2}(|x_i|^\xi|v_i|^2+
|x_{i+1}|^\xi|v_{i+1}|^2).
\end{split}
\end{equation}

Similarly,
\begin{equation}
\begin{split}
|\langle v_i,d(x_i)v_{i+1}\rangle|&\le(1+\frac{\xi}{d-1})|x_i|^\xi|v_i||v_{i+1}|\\
&\le(1+\frac{\xi}{d-1})\bigl(\frac{|x_i|}{|x_{i+1}|}\bigr)^{\xi/2}|x_i|^{\xi/2}|x_{i+1}|^{\xi/2}|v_i||v_{i+1}|\\
&\le(\frac 12+\frac{\xi}{2d-2})\bigl(\frac{|x_i|}{|x_{i+1}|}\bigr)^{\xi/2}
(|x_i|^\xi|v_i|^2+|x_{i+1}|^\xi|v_{i+1}|^2).
\end{split}
\end{equation}

Therefore, by setting $E:=\max\{\frac C2,\frac 12+\frac{\xi}{2d-2}\}$, we
can conclude our claim.
\ifcmp\qed\fi
\end{proof}

\begin{proof}(of \lemref{lem-cro2})
We just estimate $|\langle v_i,(d(v_{i,j})-d(v_{i+1,j}))v_j\rangle|$ and
the other part is estimated similarly.

Again an application of mean value theorem gives us
\begin{equation}
\begin{split}
|\langle v_i,(d(x_{i,j})-&d(x_{i+1,j}))v_j\rangle|\\&\le
\sup_{0\le r\le 1}\langle v_i,(x_i\cdot\nabla)d(x_{i+1,j}+r x_i)v_j\rangle\\
&\le\sup_{\frac{1}{2}\le |y|\le\frac{3}{2}}|\langle \hat{v_i},(\hat{x_i}\cdot\nabla)d(y))
\hat{v_j}\rangle||x_i||x_{i+1,j}|^{\xi-1}|v_i||v_j|\\
&:=C|x_i||x_{i+1,j}|^{\xi-1}|v_i||v_j|\\
&=C\bigl(\frac{|x_i|}{|x_{i+1,j}|}\bigr)^{1-\xi/2}\bigl(\frac{|x_{i+1,j}|}{|x_j|}\bigr)^{\xi/2}|x_i|^{\xi/2}|x_j|^{\xi/2}|v_i||v_j|\\
&\le\frac C2\bigl(\frac{|x_i|}{|x_{i+1,j}|}\bigr)^{1-\xi/2}\bigl(\frac{|x_{i+1,j}|}{|x_j|}\bigr)^{\xi/2}(|x_i|^\xi|v_i|^2+|x_{i+1}|^\xi|v_{i+1}|^2).
\end{split}
\end{equation}
\ifcmp\qed\fi
\end{proof}

\begin{proof}(of \lemref{lem-cro3})
First we make a split:
\begin{multline}
|\langle v_i,(d(x_{i,j})-d(x_{i+1,j})-d(x_{i,j-1})+d(x_{i+1,j-1}))v_j\rangle|
\\\le|\langle v_i,(d(x_{i,j})-d(x_{i+1,j})v_j\rangle|+
|\langle v_i,d(x_{i,j-1})v_j\rangle|+
|\langle v_i,d(x_{i+1,j-1})v_j\rangle|.
\end{multline}

Now the first term is estimated exactly as in the previous Lemma and
the latter as follows. (Actually we only estimate the second one, the
third one is handled identically).
\begin{equation}
\begin{split}
|\langle v_i,&d(x_{i,j-1})v_j\rangle|\\&\le(1+\frac{\xi}{d-1})|x_{i,j-1}|^\xi|v_i||v_j|\\
&\le(1+\frac{\xi}{d-1})\bigl(\frac{|x_{i,j-1}|}{|x_i|}\bigr)^\xi\bigl(\frac{|x_i|}{|x_j|}\bigr)^{\xi/2}|x_i|^{\xi/2}|x_j|^{\xi/2}|v_i||v_j|\\
&\le(\frac 12+\frac{\xi}{2d-2})\bigl(\frac{|x_i|+|x_{i+1,j-1}|}{|x_i|}\bigr)^\xi\bigl(\frac{|x_i|}{|x_j|}\bigr)^{\xi/2}
(|x_i|^\xi|v_i|^2+|x_j|^\xi|v_j|^2)\\
&\le(\frac 12+\frac{\xi}{2d-2})3^\xi\bigl(\frac{|x_i|}{|x_j|}\bigr)^{\xi/2}
(|x_i|^\xi|v_i|^2+|x_j|^\xi|v_j|^2).
\end{split}
\end{equation}
\ifcmp\qed\fi
\end{proof}

\begin{proof} (of \lemref{lem-cro4})
Two applications of the mean value theorem give us
\begin{equation}
\begin{split}
|\langle v_i,&(d(x_{i,j})-d(x_{i+1,j})-d(x_{i,j-1})+d(x_{i+1,j-1}))v_j\rangle|
\\&\le 
\sup_{0\le r\le 1}\langle v_i,((x_i\cdot\nabla)d(x_{i+1,j}+r x_i)-
(x_i\cdot\nabla)d(x_{i+1,j-1}+r x_i)v_j\rangle\\
&\le\sup_{0\le r,r'\le 1}\langle v_i,(x_i\cdot\nabla)(x_j\cdot\nabla)
d(x_{i+1,j-1}+r x_i+r' x_j)v_j\rangle\\
&\le\sup_{\frac{1}{3}\le |y|\le\frac{4}{3}}
|\langle \hat{v_i},(\hat{x_i}\cdot\nabla)(\hat{x_j}\cdot\nabla)d(y)\hat{v_j}
\rangle |x_i||x_j||x_{i+1,j-1}|^{\xi-2}|v_i||v_j|\\
&:=2E|x_i||x_j||x_{i+1,j-1}|^{\xi-2}|v_i||v_j|\\
&\le E\bigl(\frac{|x_i|}{|x_{i+1,j-1}|}\bigr)^{1-\xi/2}
\bigl(\frac{|x_j|}{|x_{i+1,j-1}|}\bigr)^{1-\xi/2}(|x_i|^\xi|v_i|^2+|x_j|^\xi|v_j|^2)
\end{split}
\end{equation}
\ifcmp\qed\fi
\end{proof}

\begin{proof} (of \lemref{lem-cro5})
We estimate the terms individually. The mean value theorem gives us
\begin{equation}
\begin{split}
|\langle v_i,&(d(x_{i,j})-d(x_A))v_j\rangle|\\
&\le 2^{\xi/2}C(\sum_{k\in[i,j]\setminus A}|x_k|)|x_{A}|^{\xi-1}|v_i||v_j|\\
&\le 2^{\xi/2}C\bigl(\frac{\sum_{k\in[i,j]\setminus A}|x_k|}
{|x_A|}\bigr)^{1-\xi/2}\bigl(\frac{|x_A|}{|x_j|}\bigr)^{\xi/2}.
\end{split}
\end{equation}

Since we assumed that $\sum_{k\in[i,j]\setminus A}|x_k|\le\frac 12\min\{|x_{k,l}|:k,l\in A, k\le l\}$, we have

\begin{equation}
\begin{split}
2^{\xi/2}&C\bigl(\frac{\sum_{k\in[i,j]\setminus A}|x_k|}
{|x_A|}\bigr)^{1-\xi/2}\bigl(\frac{|x_A|}{|x_j|}\bigr)^{\xi/2}\\
&\le 2^{\xi/2-1}C\bigl(\frac{\sum_{k\in[i,j]\setminus A}|x_k|}
{\min\{|x_{k,l}|:k,l\in A, k\le l\}-\sum_{k\in[i,j]\setminus A}|x_k|}\bigr)^{1-\xi/2}\bigl(\frac{|x_A|}{|x_j|}\bigr)^{\xi/2}\\
&{}\quad\cdot(|x_i|^\xi|v_i|^2+|x_j|^\xi|v_j|^2)\\
&\le C\bigl(\frac{\sum_{k\in[i,j]\setminus A}|x_k|}
{\min\{x_{k,l}:k,l\in A, k\le l\}}\bigr)^{1-\xi/2}\bigl(\frac{\sum_{k\in A}|x_k|}{|x_j|}\bigr)^{\xi/2}\\
&{}\quad\cdot(|x_i|^\xi|v_i|^2+|x_j|^\xi|v_j|^2)
\end{split}
\end{equation}
Similar estimates hold for the other terms, except when $A=\{i,j\}$, which
causes modifications to the last pair of terms. Then
\begin{equation}
\begin{split}
|\langle v_i,&d(x_{i+1,j-1})v_j\rangle|\\
&\le(1+\frac\xi{d-1})|x_{i+1,j-1}|^\xi|v_i||v_j|\\
&\le(\frac 12+\frac\xi{2d-2})\bigl(\frac{\sum_{k\in[i,j]\setminus A}|x_k|}
{|x_j|}\bigr)^{\xi/2}(|x_i|^\xi|v_i|^2+|x_j|^\xi|v_j|^2).
\end{split}
\end{equation}
\ifcmp\qed\fi
\end{proof}

\section{Proof of \proref{pro-bai}}\label{app-prc}

In order to prove \proref{pro-bai} we first need a Lemma.

\begin{apulause}\label{lem-prc}
Let $2\le l\in\N$. Then there is $C<\infty$ such that
\begin{equation}
\int_{\R^d}d^dy\,(k+|x-y|)^{2-\xi-ld}(1+|y|)^{2-\xi-d}
\le C k^{2-\xi-(l-1)d}(1+|x|)^{2-\xi-d}.
\end{equation}
\end{apulause}

\begin{proof}
We split the domain of integration into three parts and estimate these
separately:
\begin{multline}
\int_{\R^d}d^dy\,(k+|x-y|)^{2-\xi-ld}(1+|y|)^{2-\xi-d}\\
=\int_{|x-y|\le|x|/2}+\int_{|y|\le|x|/2}+
\int_{|y|,|x-y|\ge |x|/2}=:(*^1)+(*^2)+(*^3)
\end{multline}

To estimate $(*^1)$ we note that in $|x-y|\le |x|/2$ we have $|x|/2\le|y|$
which implies that in $|x-y|\le |x|/2$ we have
\begin{equation}
(1+|y|)^{2-\xi-d}\le (1+|x|/2)^{2-\xi-d}\le C_1(1+|x|)^{2-\xi-d}.
\end{equation}

Therefore
\begin{equation}
\begin{split}
(*^1)&\le C_1\int_{|x-y|\le|x|/2}d^dy\,(k+|x-y|)^{2-\xi-ld}(1+|x|)^{2-\xi-d}\\
&=C_1 k^{2-\xi-(l-1)d}(1+|x|)^{2-\xi-d}\int_{|x-y|\le|x|/(2k)}d^dy\,
(1+|x-y|)^{2-\xi-ld}\\
&\le C_1 k^{2-\xi-(l-1)d}(1+|x|)^{2-\xi-d}\int_{\R^d}d^dy\,(1+|x-y|)^{2-\xi-ld}\\
&\le C_2 k^{2-\xi-(l-1)d}(1+|x|)^{2-\xi-d}.
\end{split}
\end{equation}

We make in a similar estimate in $(*^2)$: in $|y|\le |x|/2$
we have $|x|/2\le|x-y|$ which implies that in $|y|\le |x|/2$ we have
\begin{equation}
(k+|x-y|)^{2-\xi-ld}\le (k+|x|/2)^{2-\xi-ld}\le C_3(k+|x|)^{2-\xi-ld}.
\end{equation}

Now we can compute:
\begin{equation}
\begin{split}
(*^2)&\le C_3\int_{|y|\le|x|/2}d^dy\,(k+|x|)^{2-\xi-ld}(1+|y|)^{2-\xi-d}\\
&= C_4(k+|x|)^{2-\xi-ld}\cdot
\begin{cases}
|x|^d&\text{if $|x|\le 1$ and}\\
|x|^{2-\xi}&\text{if $|x|\ge 1$}.
\end{cases}
\end{split}
\end{equation}

To treat the case $|x|\le 1$, we compute:
\begin{equation}
\begin{split}
C_4(k+|x|)^{2-\xi-ld}|x|^d&\le C_4 k^{2-\xi-(l-1)d}|x|^{-d}|x|^d\\
&\le C_5 k^{2-\xi-(l-1)d}(1+|x|)^{2-\xi-d}.
\end{split}
\end{equation}

If $|x|\ge 1$ we have
\begin{equation}
\begin{split}
C_4(k+|x|)^{2-\xi-ld}|x|^{2-\xi}&\le C_4 k^{2-\xi-(l-1)d}|x|^{-d}|x|^{2-\xi}\\
&=C_4 k^{2-\xi-(l-1)d}|x|^{2-\xi-d}.
\end{split}
\end{equation}

Finally, we handle $(*^3)$. When $|y|,|x-y|\ge|x|/2$, we have
$|y|/3\le|x-y|$. Since this might not be obvious, we compute: Since
$B(x,|x|/2)\subseteq B(0,3|x|/2)$, we have
\begin{equation}
\begin{split}
|y|/3&= |x|/2+\frac 13 d( y, B(0,3|x|/2))\\
&\le |x|/2+d(y,B(0,3|x|/2)\\
&\le |x|/2+d(y,B(x,|x|/2)\\
&=|x-y|.
\end{split}
\end{equation}

Therefore, when $|y|,|x-y|\ge |x|/2$, we have
\begin{equation}
(k+|x-y|)^{2-\xi-ld}(1+|y|)^{2-\xi-d}\le C_6 (k+|y|)^{2-\xi-ld}(1+|y|)^{2-\xi-d}
\end{equation}
and thus
\begin{equation}
(*^3)\le C_6\int_{|y|\ge|x|/2}d^dy\,(k+|y|)^{2-\xi-ld}(1+|y|)^{2-\xi-d}=:(*^4)
\end{equation}

We split the analysis of $(*^4)$ into two subcases: $|x|\ge 2$ and $|x|\le 2$.

If $|x|\ge 2$, then we have
\begin{equation}
\begin{split}
(*^4)&\le C_6\int_{|y|\ge|x|/2}\,(k+|y|)^{2-\xi-ld}|y|^{2-\xi-d}\\
&=C_7 k^{2(2-\xi)-ld}\int_{|y|\ge |x|/(2k)}d^dy\,(1+|y|)^{2-\xi-ld}|y|^{2-\xi-d}\\
&=C_8 k^{2(2-\xi)-ld}(1+|x|/k)^{2(2-\xi)-ld}=:(*^5)
\end{split}
\end{equation}

If $|x|\le k$, then 
\begin{equation}
(*^5)=C_8 k^{2(2-\xi)-ld}\le C_9 k^{2-\xi-(l-1)d}(1+|x|)^{2-\xi-d}.
\end{equation}

On the other hand, if $k\le |x|$, then
\begin{equation}
(*^5)=C_8 |x|^{2(2-\xi)-ld}\le C_{10} k^{2-\xi-(l-1)d}(1+|x|)^{2-\xi-d}.
\end{equation}

If instead of $|x|\ge 2$ we have $|x|\le 2$ in $(*^4)$, we compute
\begin{equation}
\begin{split}
(*^4)&\le C_6\int_{\R^d}d^dy\,(k+|y|)^{2-\xi-ld}(1+|y|)^{2-\xi-d}\\
&\le C_{11}\int_{|y|\le 1}d^dy\,(k+|y|)^{2-\xi-ld}+C_{11}\int_{|y|\ge 1}(k+|y|)^{2-\xi-ld}|y|^{2-\xi-d}\\
&\le C_{12} k^{2-\xi-(l-1)d}\int_{\R^d}d^dy\,(1+|y|)^{2-\xi-ld}+C_{12} k^{2-\xi-(l-1)d}\\
&\le C_{13} k^{2-\xi-(l-1)d}(1+|x|)^{2-\xi-d}.
\end{split}
\end{equation}
\ifcmp\qed\fi
\end{proof}

\begin{proof} (of \proref{pro-bai}.)
Without loss of generality, we may assume that $\chi$ is the characteristic
function of the unit ball. First we integrate $y_l$ out:

Write $k:=\sum_{i=1}^{l-1}|x_i-y_i|$. Now we have
\begin{equation}
\begin{split}
\int_{y_l\in B(0,1)}&d^dy_l\,(k+|x_l-y_l|)^{2-\xi-ld}\\
&\le C_1
\begin{cases}
(k+|x_l|)^{2-\xi-ld}&\text{if $|x_l|\ge 2$,}\\
k^{2-\xi-(l-1)d}&\text{if $|x_l|\le 2$ and $k\le 1$ and}\\
k^{2-\xi-ld}&\text{if $|x_l|\le 2$ and $k\ge 1$}.
\end{cases}
\end{split}
\end{equation}

The first case, i.e. $|x_l|\ge 2$, is computed by a repeated application of
Lemma \ref{lem-prc}:

\begin{equation}
\begin{split}
\int_{\R^{(l-1)d}}\prod_{i=1}^{l-1}d^d&y_i\,(|x_l|+\sum_{i=1}^{l-1}|x_i-y_i|)^{2-\xi-ld}\prod_{i=1}^{l-1}(1+|y_i|)^{2-\xi-d}\\
&\le C_2(1+|x_{l-1}|)^{2-\xi-d}\int_{\R^{(l-1)d}}\prod_{i=1}^{l-1}d^dy_i\cdot\\
&\cdot(|x_l|+\sum_{i=1}^{l-2}|x_i-y_i|)^{2-\xi-(l-1)d}\prod_{i=1}^{l-2}(1+|y_i|)^{2-\xi-d}\\
&\le ...\\
&\le C_{l}\prod_{i=2}^{l-1}(1+|x_i|)^{2-\xi-d}\int_{\R^{d}}d^dy_1\cdot\\
&\cdot(|x_l|+|x_1-y_1|)^{2-\xi-2d}(1+|y_1|)^{2-\xi-d}\\
&\le C_{l+1}\prod_{i=1}^l(1+|x_i|)^{2-\xi-d}.
\end{split}
\end{equation}

In the second case, i.e. $|x_l|\le 2$ and $k\le 1$, we get
\begin{equation}
\begin{split}
\int_{k\le 1}&\prod_{i=1}^{l-1} d^dy_i\,k^{2-\xi-(l-1)d}
\prod_{i=1}^{l-1}(1+|y_i|)^{2-\xi-d}\\
&\le\sup_{k\le 1}\prod_{i=1}^{l-1}(1+|y_i|)^{2-\xi-d}\int_{k\le 1}\prod_{i=1}^{l-1} d^dy_i\,k^{2-\xi-(l-1)d}\\
&\le C'\prod_{i=1}^{l-1}(1+|x_i|)^{2-\xi-d}.
\end{split}
\end{equation}

The third case, i.e. $|x_l|\le 2$ and $k\ge 1$, uses the following trick:
\begin{equation}
\begin{split}
\int_{k\ge 1}&\prod_{i=1}^{l-1}d^dy_1\,k^{2-\xi-ld}
\prod_{i=1}^{l-1}(1+|y_i|)^{2-\xi-d}\\
&\le C'_2 \int_{\R^{(l-1)d}}\prod_{i=1}^{l-1}d^dy_i\,(1+k)^{2-\xi-ld}
\prod_{i=1}^{l-1}(1+|y_i|)^{2-\xi-d}=(*)
\end{split}
\end{equation}

Now repeating the computation of the first case, we get:
\begin{equation}
(*)\le C'_3\prod_{i=1}^{l-1}(1+|x_i|)^{2-\xi-d}\le C'_4\prod_{i=1}^l(1+|x_i|)^{2-\xi-d}.
\end{equation}
\ifcmp\qed\fi
\end{proof}

\ifcmp 
\begin{acknowledgements} 
\acknowledgementstext 
\end{acknowledgements} 
\fi 

\bibliographystyle{amsplain}

\end{document}